\documentclass[%
 reprint,
superscriptaddress,
 amsmath,amssymb,
 aps,
pre,
]{revtex4-2}

\usepackage{dcolumn}
\usepackage{bm}
\usepackage{comment} 
\usepackage{hyperref}
\usepackage{amssymb,amsmath,graphicx,array,color,amsthm}
\usepackage{caption}
\usepackage[skip=-5pt]{subcaption}
\usepackage{amscd}

\usepackage[normalem]{ulem}

\def\kk{{\bf k}}
\def \BE {\begin{equation}}
\def \EE {\end{equation}}
\def \BEA {\begin{eqnarray}}
\def \EEA {\end{eqnarray}}

\newcommand*{\ADD}[1]{{\color{magenta} #1}}

\begin{document}
\preprint{AIP/123-QED}

\title{Testing wave turbulence theory for Gross-Pitaevskii system }

\author{Ying Zhu}
\email{yzhu@unice.fr}
\affiliation{Universit\'{e} C\^{o}te d'Azur, CNRS, Institut de Physique de Nice {\color{black}(INPHYNI)}, Parc Valrose, 06108 Nice, France}
\author{Boris Semisalov}
\affiliation{Universit\'{e} C\^{o}te d'Azur, Observatoire de la C\^{o}te d'Azur, CNRS, Laboratoire Lagrange, Boulevard de l'Observatoire CS 34229 -- F 06304 Nice Cedex 4, France}
\affiliation{Novosibirsk State University, 1 Pirogova street, 630090 Novosibirsk, Russia}
\affiliation{Sobolev Institute of Mathematics SB RAS, 4 Academician Koptyug Avenue, 630090 Novosibirsk, Russia}
\author{Giorgio Krstulovic}
\affiliation{Universit\'{e} C\^{o}te d'Azur, Observatoire de la C\^{o}te d'Azur, CNRS, Laboratoire Lagrange, Boulevard de l'Observatoire CS 34229 -- F 06304 Nice Cedex 4, France}
\author{Sergey Nazarenko}
\affiliation{Universit\'{e} C\^{o}te d'Azur, CNRS, Institut de Physique de Nice - INPHYNI, Parc Valrose, 06108 Nice, France}

\begin{abstract}
  We test the predictions of the theory of weak wave turbulence by performing numerical simulations of the Gross-Pitaevskii equation (GPE)  and the associated wave-kinetic equation (WKE). We consider an initial state localized in Fourier space, and we confront the solutions of the WKE obtained numerically with GPE data for both the wave-action spectrum and the probability density functions (PDFs) of the Fourier mode intensities. We find that the temporal evolution of \textcolor{black}{the GPE} data is accurately predicted by the WKE, with no adjustable parameters, for about two nonlinear kinetic times.
  Qualitative agreement between the GPE and the WKE persists also for longer times with some quantitative deviations that may be attributed to the \textcolor{black}{combination of} breakdown of the theoretical assumptions underlying the WKE \textcolor{black}{as well as numerical issues}. Furthermore, we study how the wave statistics evolves toward Gaussianity in a time scale of the order of the kinetic time. The excellent agreement between direct numerical simulations of the GPE and the WKE provides a new and solid ground to the theory of weak wave turbulence.
\end{abstract} 

 \maketitle  

\section{Introduction }

Wave Turbulence (WT) is a state of continuous medium characterized by presence of random mutually interacting waves with a broadband spectrum \textcolor{black}{ \cite{nazarenko2011wave,ZLF,dyachenko1992optical}}. Weak WT theory  is a mathematical framework  describing the statistical behavior of WT dominated by weakly nonlinear waves.
 The main object in this theory is the wave-action spectrum which is the second-order moment of the wave amplitude and which evolves according to the so-called wave-kinetic equation (WKE). Special attention in past literature  was given to studies of stationary scaling solutions of this equation which are \textcolor{black}{ similar} to the Kolmogorov spectrum of hydrodynamic turbulence, the so-called Kolmogorov--Zakharov (KZ) spectra (see e.g. {\color{black}\cite{proment2009energy,proment2012sustained, navon2016emergence, navon2019synthetic}}). However, 
 nonstationary WT is also interesting because it is often characterized by mathematically nontrivial solutions exhibiting self-similar asymptotic \textcolor{black}{behaviour} \cite{semikoz1995kinetics,lacaze2001dynamical,connaughton2004kinetic,BellNaz,Bell_2018,SemGreMedNaz}.
 In general, nonstationary WT allows more subtle tests of the weak WT theory
 than stationary setups, which is why we will consider it in the present paper.
 Further, the weak WT theory   can be also extended to describing the higher-order moments and even to the full joint probability density function (PDF) of the wave intensities \cite{nazarenko2011wave,lvov_2004,choi_2004,choi_2005,choi_2005b}. 
 The fundamental concepts of WT were laid out 
in the classical book \cite{ZLF}.
An introduction to WT as well as a summary of recent developments in this area can be found in book  \cite{nazarenko2011wave}.  Recent book \cite{shrira2013advances} contains a collection of reviews about recent experiments in WT. 

Over the past few years, a significant push was made in the direction of rigorous mathematical justification of the weak WT theory. The state of the art result is a claim made in \cite{hani} that the WKE~\eqref{KE0} derived from the 3D Gross-Pitaevskii equation (GPE)~\eqref{GPE} truthfully predicts evolution of the wave spectrum (under certain requirements on the initial conditions) up to a time $\delta \cdot T_{kin} $ where $ T_{kin} $ is the characteristic evolution time of the WKE and $\delta \ll 1 $. This result is significant from the mathematical point of view because $\delta$ is independent of the (vanishing) nonlinearity strength and the (expanding to infinity) physical size of the system. As such this result is the wave analog of the famous Lanford's theorem on the Boltzmann kinetic equation from colliding particle systems. However, 
this result is insufficient for physical applications, in the majority of which one is interested in evolution at times $\gtrsim T_{kin} $ when the spectrum has significantly evolved away from its initial shape, whereas at time $\delta \cdot T_{kin} $ it is still very close to the initial one.

Thus, the main motivation of the present study is to test weak WT theory  at the times $\gtrsim T_{kin} $ by juxtaposition of the spectra arising from the direct numerical simulations (DNS) of the 3D GPE~\eqref{GPE} and from the simulations of the WKE~\eqref{KE0}. Further,  we explore numerically  the evolution of the wave-intensity PDFs and the statistical quantities associated with it, e.g. the cumulants. We show that both the spectra and the PDFs for the WKE and the GPE agree with each other very well for times $\gtrsim T_{kin} $, and we characterize the departures at later times. 

The first comparisons  for the spectra, PDFs and cumulants obtained from a DNS and simulations of the weak WT  closure in \cite{Tanaka} for a model 2D three-wave system  (using simulations up to  $ \sim 0.3 T_{kin}$).
 They arrived at favourable conclusions for validity of the weak WT theory.
 The system considered in the present work is 3D and four-wave, and our simulations extend to about $ 10 T_{kin}$.
 In addition, in the present work we take advantage of the analytical expression for the PDF in terms of the wave spectrum which was obtained in \cite{Choi_2017} and, therefore, was known at the time of writing paper \cite{Tanaka}.

Further, a   comparison of the evolving spectra arising in the numerics of WKE and the original dynamic model was also done in~\cite{PhysRevLett.99.164501,KOROTKEVICH} for the deep water surface gravity waves. Their study was oriented on the practical problem of the sea swell modeling, with comparisons made on a qualitative level of visual comparing the plots. Further modifications by introducing dissipation and forcing were explored with a view to make the models more realistic for practical modeling of the sea waves {\color{black}\cite{korotkevich2019dissipation}}. In contrast, our study  is aimed at the more fundamental aspects of the \textcolor{black}{weak WT theory} validity rather than a particular real life wave phenomenon. In order to make the comparisons be like-for-like, we do not introduce any extra terms to improve the modeling. We introduce global measures characterizing the spectra, such as the energy and the wave-action centroids, to make our comparisons more  quantitative.  Finally, note that the recent work \cite{banks2021direct} compares the evolution of the one-dimensional defocusing quintic nonlinear Schroedinger equation with the evolution of the associated WKE, finding a good agreement when some scaling relations between nonlinearity strength and system size are fulfilled.  

The present paper is organized as follows. In Section \ref{sec:theory} we introduce the GPE and the setting we consider in this work. We provide then a short introduction to the theory of weak wave turbulence and present the wave-kinetic equation. The dynamics of the probability distribution functions of the wave-action spectra is also discussed. In Section~\ref{sec:numerics}, we explain our numerical methods for integrating the GPE and the WKE. Then, in Section~\ref{sec:NumResults} we present our numerical results where we directly compare numerical simulations of the GPE and the WKE. Finally, in Section~\ref{sec:conclu} we present our conclusions. We took out the details on derivation of the WKE in isotropic homogeneous setup to the Appendix~\ref{KEder} and discuss the most important issues of the new numerical algorithm used for solving WKE in Appendix~\ref{Numerics}.  

\section{Theoretical background} \label{sec:theory}

\subsection{Gross-Pitaevskii equation}

The dimensionless Gross-Pitaevskii Equation (GPE) in a 3D physical space ${\bm r} =\{x,y,z\}$  for the complex wave function $\psi({\bm r},t)$ is
  \begin{align}\begin{split}
  i \frac{\partial\psi ({\bm r},t)  }{\partial t}=&\Big [ -\nabla^{2}  +\left|\psi({\bm r},t) \right|^{2}  \Big ]\psi({\bm r},t) \ .
  \label{GPE}
 \end{split} \end{align} 
\textcolor{black}{ GPE is also known as defocusing Nonlinear Schr\"odinger equation (NLSE).}
 This equation describes dynamics of Bose-Einstein condensates (BEC)~\cite{pitaevskii2003bose}, nonlinear light~\cite{dyachenko1992optical}
 and other important physical systems, and as such it is a universal master equation that 
 allows to study fundamental nonlinear phenomena including the nonlinear wave interactions.
Consider the nonlinear wave system described by the GPE \eqref{GPE} in the $L$-periodic box of volume  $V=L^3$. From GPE \eqref{GPE} we have the following Hamiltoninan equations for the Fourier coefficients of the wave function,
\begin{equation}
\begin{aligned}
&i\dot{a_{\bf k}} = \frac{\partial \mathcal{H}}{\partial {a}_{\bf k}^*},\\
&\mathcal{H}=\sum_{\bf k} \omega_{\bf k} |{a}_{\bf k}|^2 + \frac{1}{2}\sum_{{\bf k}_1, {\bf k}_2,{\bf k}_3,{\bf k}_4} 
{a}_{{\bf k}_1}^* {a}_{{\bf k}_2}^* a_{{\bf k}_3} a_{{\bf k}_4}\delta_{12}^{34},
\end{aligned}\label{4wH}
\end{equation}
where ${\bf k}, {\bf k}_1, {\bf k}_2,{\bf k}_3,{\bf k}_4 \in \textcolor{black}{\frac  {2\pi} L} \mathbb{Z}^3 $ are the wave vectors, $\omega_{\bf k} =k^2$ is the frequency of \textcolor{black}{the linear wave with wave vector} $\bf k$, $k=|{\bf k}|$, and $a_{\bf k}\in \mathbb{C}$ is the wave-action variable:
\begin{equation}
\label{ak}
a_{\bf k} = \hat{\psi}_{\bf k} = \frac{1}{V}\int\psi({\bm r},t)e^{-i{\bf k}\cdot\bf{r}}d{\bf r},
\end{equation}
$\delta_{12}^{34}=\delta(\bm{k}_1+\bm{k}_2-\bm{k}_3-\bm{k}_4)$, and the integral is taken over all the possible discrete values of $\bm{k}_1$, $\bm{k}_2$, $\bm{k}_3$ and $\bm{k}_4$.

The GPE \eqref{GPE} conserves  the mean density of particles, 
\begin{equation}
N =\frac{1}{V}\int |\psi(\mathbf{x}, t)|^2 d\mathbf{x} \,
\label{eq:cons-N}
\end{equation}
and the mean density of  energy,
\begin{equation}
H=\frac{1}{V}\int \left[ |\nabla \psi(\mathbf{x}, t)|^2 + \frac{1}{2}|\psi(\mathbf{x}, t)|^4 \right] d\mathbf{x} \, .
\label{eq:cons-H}
\end{equation}
The total energy density  can be split into two parts -- the energy density of the linear-dynamics $H_2=\frac{1}{V}\int  |\nabla \psi(\mathbf{x}, t)|^2  d\mathbf{x}$, and the nonlinear-dynamics energy $H_4=\frac{1}{V}\int  \frac{1}{2}|\psi(\mathbf{x}, t)|^4  d\mathbf{x}$.
In a wave turbulent state, the total energy and the density of particles cascade in opposite directions {\color{black}\cite{nazarenko2011wave,dyachenko1992optical}, which is similar to the dual cascade behavior in 2D hydrodynamic turbulence \cite{fjortoft1953changes}}.

An important quantity, which characterizes the system, is the healing length, $\xi=\frac{1}{\sqrt{N}}$, which refers to an average  scale at which the nonlinear term becomes comparable with the linear one. The characteristic wave number corresponding to $\xi$ is defined as $k_{\xi}=\frac{1}{\xi}$.

Also, we define the condensate fraction as 
\begin{equation}
C_0=\frac{|\hat{\psi}_{\bm{0}}|^2}{N}.
\label{eq:C0}
\end{equation}
$C_0=0$ means that there is no homogeneous condensate in the field, and $C_0=1$ represents a complete condensation.
When the condensate fraction is small or absent, and the waves are weak, the system described by GPE can be modelled within the \textcolor{black}{weak WT }  approach leading to the four-wave WKE \cite{nazarenko2011wave} introduced in the next subsection.

\subsection{Wave-kinetic equation description}

The analysis of the GPE in the case of weak nonlinearity leads to the integro-differential WKE describing the evolution of the wave-action spectrum 
\begin{equation}
n_{\bf k}(t)=n({\bf k},t) = \biggl(\frac{L}{2\pi}\biggr)^3\langle |a_{\bf k}|^2 \rangle,
\end{equation}
where the brackets denote {\color{black}averaging
over the ensemble of initial conditions.}

To make all further considerations consistent with the \textcolor{black}{weak WT theory}, let us briefly
summarise the basic steps of derivation of WKE from GPE \cite{nazarenko2011wave}:

\begin{itemize}
\item Consider GPE in the Fourier space. 

\item Set a small parameter \textcolor{black}{$0<\varepsilon\ll 1$}, which is the measure of nonlinearity, and consider GPE on the intermediate time scale $T_I$ between the linear and nonlinear time scales:
$$
\frac{2\pi}{k^2} \text{(linear scale)}\ll 
T_I\sim\frac{2\pi}{\varepsilon k^2}
\ll \frac{2\pi}{\varepsilon^2 k^2}\text{(nonlinear scale)}.
$$
\item Introduce the interaction representation variable, 
\begin{equation}
\hat{b}_{\bf k}(t)=\frac{1}{\sqrt{\varepsilon}}\hat{\psi}_{\bf k}\exp(i\tilde{\omega}_{\bf k}t),~\tilde{\omega}_{\bf k}=k^2+2N.
\label{eq:ST}
\end{equation}
Here, the second term in $\tilde{\omega}_{\bf k}$ is the so-called nonlinear frequency shift: it is the leading order nonlinear effect, but it does not lead to energy exchanges between the Fourier modes.

\item assume  that  initially the  amplitudes $|\hat{b}_k|$ and the phase factors $\hat{b}_{\bf k}/|\hat{b}_{\bf k}|$ are statistically independent random variables, and the phase factors are uniformly distributed on the unit circle in the complex plane. 
\textcolor{black}{This was called the
Random Phase and Amplitude (RPA) statistics in \cite{choi_2004,choi_2005, choi_2005b} which is a change to the standard meaning of this acronym ``random phase approximation" made in order to emphasize that the amplitude randomness is  essential for the WT closure and the fact that there is no approximation is involved in taking initial conditions of this type.}
\item Make the asymptotic expansion of $\hat{b}_{\bf k}(T_I)$ with respect to the small parameter $\varepsilon$, substitute it into GPE and find the averaged coefficients with the aid of the RPA property.

\item Approximate the time derivative of the wave-action spectrum by difference quotient, express it through the obtained averaged coefficients and pass first to the large box limit $L\to\infty$ (which makes the $k$-space continuous). Then pass to the limit $T_I\to\infty$.
\end{itemize}

As the result, one obtains the following WKE for the four-wave interaction ${\bf k}, {\bf k}_1\to {\bf k}_2, {\bf k}_3$ \cite{ZAKHAROV1985285}:
\begin{widetext}
\begin{equation}\label{KE0}
\frac{d}{dt}n_{\bf k} = 4\pi\int\delta({{\bf k} + {\bf k}_1 -{\bf k}_2 -{\bf k}_3} )\delta(\omega_{{\bf k}} + \omega_{{\bf k}_1} - \omega_{{\bf k}_2}
-\omega_{{\bf k}_3})\big[n_{{\bf k}_1}n_{{\bf k}_2}n_{{\bf k}_3}+n_{{\bf k}}(n_{{\bf k}_2}n_{{\bf k}_3}-n_{{\bf k}_1}n_{{\bf k}_3}-n_{{\bf k}_1}n_{{\bf k}_2})\big]d{\bf k}_1d{\bf k}_2d{\bf k}_3,
\end{equation} 
\end{widetext}
where $\delta$ is the Dirac $\delta$-function, and wavevectors are now continuous (${\bf k}_i\in\mathbb{R}^3$). 



The collision integral in the right-hand-side (RHS) of \eqref{KE0} is taken over a 9-dimensional space. However, we shall consider the situation when the wave fields are isotropic. In this situation, averaging over directions  reduces considerably the dimension of integration. Such simplification gives the possibility to study complex non-stationary self-similar solutions of the second kind, see for instance \cite{semikoz1995kinetics,semikoz1997condensation,lacaze2001dynamical,connaughton2004kinetic,SemGreMedNaz}. Performing the angle averaging leads to a new collision kernel. Previous results presented in the literature reported disparate pre-factors of the kernel \cite{semikoz1995kinetics, ZLF, DMPS}. The value of this pre-factor is crucial for a quantitative comparison of the evolution of GPE and WKE solutions. In Appendix~\ref{KEder} we provide a careful derivation, where we have corrected previous errors.

Assuming the isotropy, we consider the wave-action spectrum that depends only on $k=|{\bf k}|$ and pass to the frequency variable: $n_{\bf k}(t) = n_{\omega}(t)=n(\omega,t)$, where $\omega=k^2$ is the new variable. The isotropic WKE is
\begin{eqnarray}\label{E01}
\frac{d}{dt}n_{\omega} = \frac{4\pi^3}{\sqrt{\omega}}\int S(\omega,\omega_1,\omega_2,\omega_3)\delta_{1\omega}^{23}~
n_{\omega}n_{1}n_{2}n_{3} \\
\left(n_{\omega}^{-1} + n_{1}^{-1} - n_{2}^{-1} - n_{3}^{-1}\right)d\omega_1d\omega_2d\omega_3. \nonumber
\end{eqnarray} 
where $\delta_{1\omega}^{23}=\delta(\omega+\omega_1-\omega_2-\omega_3)$. The integral in the RHS of (\ref{E01}) is taken over all positive values of $\omega_{1}$, $\omega_{2}$, $\omega_{3}$. 
The kernel of the integral is
\begin{equation}
S(\omega,\omega_1,\omega_2,\omega_3) = \min\left(\sqrt{\omega},\sqrt{\omega_1},\sqrt{\omega_2},\sqrt{\omega_3}\right).
\end{equation}

Making use of $\delta_{1\omega}^{23}$, and defining $n_c(t) = n(\omega_2+\omega_3-\omega,t)$, $S_{\omega}^{23} = S(\omega,\omega_2+\omega_3-\omega, \omega_2,\omega_3)$ and 
$\Delta_{\omega}=\{(\omega_2,\omega_3):\omega_2,\omega_3\geq 0,~\omega_2+\omega_3\geq \omega\}$, one can do further simplifications of the WKE. Finally, it reduces to
\begin{equation}\label{E1}
  \frac{d}{dt}n_{\omega} =  \eta_{\omega}(t) - n_{\omega}(t)\gamma_{\omega}(t),
  \end{equation} 
where 
\begin{eqnarray}
\label{eta}
    \eta_{\omega}(t) &=& \frac{4\pi^3}{\sqrt{\omega}} \int\limits_{\Delta_{\omega}} S_{\omega}^{23} n_cn_2n_3d\omega_2d\omega_3,\\
    \gamma_{\omega}(t) &=& \frac{4\pi^3}{\sqrt{\omega}} \int\limits_{\Delta_{\omega}} S_{\omega}^{23} \big[n_c(n_2+n_3)-n_2n_3\big]d\omega_2d\omega_3. \,
\label{gam}
\end{eqnarray}

Equation \eqref{E1}  conserves the density of particles,
\begin{equation}
\label{pa}
N =  2\pi \int\limits_0^\infty  \omega^{1/2} \, n_{\omega} \, d \omega ,
\end{equation}
and the density of linear-dynamics energy, 
\begin{equation}
\label{en}
H=H_2 = 2 \pi \int\limits_0^\infty   \omega^{3/2}  \, n_{\omega} \, d \omega.
\end{equation}
Note that in the \textcolor{black}{weak WT theory}  the nonlinear-dynamics energy drops out from the invariant (\ref{eq:cons-H}). This is natural because by construction it is much smaller than the linear-dynamics energy.

{\color{black}Because the system we study is approximatelly statistically isotropic, it is natural to introduce the spherically-integrated (radial) spectra of wave-action and energy defined as follows:}
\begin{eqnarray}
\label{eq:nk1D} 
{\color{black}n^{\text{rad}}}(k,t)&=&\int n_{\bf k}(t) k^2d\Omega,\\
\label{eq:Ek1D} 
{\color{black}E^{\text{rad}}}(k,t)&=&\int n_{\bf k}(t) k^4d\Omega = k^2{\color{black}n^{\text{rad}}}(k,t),
\end{eqnarray}
where $d\Omega$ is the surface element of the unit sphere in the 3D ${\bf k}$-space. 

{\color{black}Note that the $n^{\text{rad}}(k,t)$ and $E^{\text{rad}}(k,t)$ indicate the distributions of particles and energy over $k$; respectively, we have the relations}
$$N=\int\limits_0^{\infty} {\color{black}n^{\text{rad}}}(k,t) d k~\text{ and }~H_2 =\int\limits_0^{\infty} {\color{black}E^{\text{rad}}}(k,t) d k.$$

\subsection{Wave turbulence description beyond the spectrum}

It has been widely believed that the statistics of random weakly nonlinear wave systems
is close to being Gaussian. Derivation of the evolution equation for the probability distribution function (PDF) of the wave intensities presented in \cite{choi_2005} has made it possible to examine  this belief.
It was shown  in \cite{choi_2005} that this equation indeed has a stationary solution corresponding to the Gaussian state, but it was also noted that the typical evolution time of the PDF is the same as the one for the spectrum. Thus, for non-stationary wave systems one can expect significant deviations from the Gaussianity if the initial wave distributions are non-Gaussian. Note that non-Gaussian (usually deterministic) initial conditions for the wave intensity are typical in numerical simulations in WT. Also, there is no reason to believe that initial waves excited in natural conditions, e.g. sea waves excited by wind, should be Gaussian. Therefore, {\color{black}the} study of evolution of the wave statistics is important for both understanding of fundamental nonlinear processes and for the practical predictions such as e.g. wave weather forecast.

In the present paper we shall use the full general solution for the PDF equation derived in \cite{Choi_2017} and also the expressions for the moments and cumulants of PDF from \cite{lvov_2004}. Here, we briefly summarize the results of references~\cite{lvov_2004,choi_2005} and~\cite{Choi_2017}.

Let us consider the PDF ${\mathcal P}(s_{\bf k},t)$
of the wave intensity $J_{\bf k}=
|a_{\bf k}|^2$ defined in a standard way, namely, the probability for
$J_{\bf k}$ to be in the range from $s_{\bf k}$ to $s_{\bf k} +ds_{\bf k}$ is
${\mathcal P}(s_{\bf k},t) d s_{\bf k}$. In symbolic form,
\begin{equation}
{\mathcal P}(s_{\bf k},t) = \langle \delta (s_{\bf k}-J_{\bf k}) \rangle.\label{pasdelta}
\end{equation}

Under the same assumptions of RPA and weak nonlinearity as the ones used for WKE derivation,
 the following evolution equation 
 for  ${\mathcal P}(s_{\bf k},t)$ was derived in \cite{choi_2005},
\begin{equation}
\frac{\partial {\mathcal P}(s_{\bf k},t)}{\partial t} +\frac{\partial F}{\partial s_{\bf k}}=0,\label{main}
\end{equation}
where
\begin{equation}
F = -s_{\bf k} \Big(\gamma_{\bf k} {\mathcal P} +\eta_{\bf k} \frac{\partial {\mathcal P} }{\partial s_{\bf k}}\Big)
\end{equation}
and, for the four-wave systems arising from the GPE,
\begin{widetext}
\begin{eqnarray}
\eta_{\bf k}(t) &=& 4\pi \int
\delta({{\bf k} + {\bf k}_1 -{\bf k}_2 -{\bf k}_3} )\delta(\omega_{{\bf k}} + \omega_{{\bf k}_1} - \omega_{{\bf k}_2}
-\omega_{{\bf k}_3}) n_{{\bf k}_1} n_{{\bf k}_2} n_{{\bf k}_3} d\kk_1d\kk_2d\kk_3, \\
\gamma_{\bf k}(t) &=& 8\pi \int
\delta({{\bf k} + {\bf k}_1 -{\bf k}_2 -{\bf k}_3} )\delta(\omega_{{\bf k}} + \omega_{{\bf k}_1} - \omega_{{\bf k}_2}
-\omega_{{\bf k}_3}) \Big[n_{{\bf k}_1}(n_{{\bf k}_2}+n_{{\bf k}_3})-n_{{\bf k}_2}n_{{\bf k}_3}\Big]d\kk_1d\kk_2d\kk_3.
\label{gam_4wave}
\end{eqnarray}
\end{widetext}
Note that in terms of $\eta_{\bf k}$ and $\gamma_{\bf k}$, the WKE \eqref{KE0} reads
$\frac{d}{dt}n_{\bf k} = \eta_{\bf k} -  \gamma_{\bf k} n_{\bf k}$.
In the isotropic case, $\eta_{\bf k} = \eta_{\omega} $ and $\gamma_{\bf k}= \gamma_{\omega}$
as defined in \eqref{eta} and \eqref{gam}.

It is important to realize that generally the PDF evolves at the same time scale $\tau_{kin}$ than the spectrum,
namely 
$\tau_{kin} \sim \min (1/\gamma, n/\eta$). 
Exceptional are the cases, when either the spectrum is close to a stationary 
one (e.g. KZ spectrum) or the cases the PDF is  close to the stationary one (corresponding to the Gaussian wave fields). In the former case the spectrum evolution time becomes very large, but not necessarily the PDF evolution time, and in the latter case -- vice versa.

Laplace transform of equation \eqref{main} converts it into a first-order  partial differential equation (PDE), which can be solved by the method of characteristics. Moreover,  this PDE is linear  with respect to ${\mathcal P}$ if we consider spectrum $n_{\bf k}(t)$ given (i.e. found by solving  the WKE first). These facts were used in \cite{Choi_2017}, and the general time-dependent solution of the equation (\ref{main}) for PDF is obtained there for arbitrary initial statistics.

The solution is constructed at a fixed wave number, so we shall drop the subscripts $\bf k$ for simplicity. 
Let us introduce the Green's function $\mathcal{P}_J(s,t)$, i.e. the solution evolving from 
the initial condition $\mathcal{P}(s,0) = \delta(s - J)$, which corresponds to the 
deterministic initial wave intensity. The general solution with an arbitrary initial condition $\mathcal{P}(s,t)$ is thus given by
\begin{equation}
\mathcal{P}(s,t) =\int_0^\infty \mathcal{P}(J,0) \mathcal{P}_J(s,t) dJ. \label{solnPgen}
\end{equation}
Using the Laplace transform and the method of characteristics, the following solution 
was obtained in \cite{Choi_2017},
\begin{eqnarray}
\label{solnPst1}
\mathcal{P}_J(s,t) 
&=& \frac{e^{-\frac{s}{\tilde n} - a \tilde n}}{2\pi i {\tilde n}}
\lim_{T\to +\infty} \int^{T+i\infty}_{T-i\infty} 
\frac{ e^{s \rho +\frac a \rho} }\rho d \rho\\
&=&
\frac 1 { {\tilde n}} {e^{-\frac{s}{\tilde n} - a \tilde n}}
I_0(2\sqrt{as}),\nonumber
\end{eqnarray}
where 
\begin{equation}
\label{nPDF}
\tilde n = n(t)-J e^{-\int^t_0 \gamma(t')dt'},
\end{equation}
$n(0) =J$, $a = \frac{J}{ {\tilde n}^2} e^{-\int^t_0 \gamma(t')dt'}$
 and $I_0(x) $ is the zeroth modified Bessel function of the first kind.
 
 This is a general solution for an arbitrary initial PDF, but we emphasize
 that $n(t)$ is considered  given. 
 Thus, finding the PDF is a two-step process: first, one has to find $n_{\bf k}(t)$ by solving the WKE (usually numerically), and second, substitute the result into the above formula for the analytical solution.
 
 Note that for Gaussian wave fields the PDF of the wave intensities is $\mathcal{P}_G = \frac 1 n e^{-s/n}$.
 As shown in \cite{Choi_2017}, solution \eqref{solnPgen}, \eqref{solnPst1} implies that
 (A)~the fields, which are Gaussian initially, will remain Gaussian for all time, and
(B)~at each fixed $s$ wave turbulence asymptotically becomes Gaussian if
  \begin{equation}
\lim_{t \to \infty} \frac {n(0) e^{-\int_0^t {\gamma(t')}dt' }}{n(t)} = 0.
\label{cond}
   \end{equation}
Indeed, taking into account that $I_0(0)=1$, we recover that $\mathcal{P}_J \to \mathcal{P}_G = \frac 1 n e^{-s/n}$
as $t \to \infty$, if condition (\ref{cond}) is satisfied provided that  $as \ll 1$.
Notice also that $a(t) \to 0$.

It is interesting that the rate of convergence to Gaussianity is greater at those $k$'s where the initial spectrum is smaller. Indeed, in the limit $J=n(0) \to 0$ we have $\tilde n \to n$, and $a\to 0$, so that $\mathcal{P}_J \to \mathcal{P}_G = \frac 1 n e^{-s/n}$.

On the other hand, for  large $t$ and $as \gg 1$,
taking into account that $I_0(x) \xrightarrow{x\rightarrow \infty} \frac{e^x}{\sqrt{2\pi x}}$, we have
\begin{equation}
\mathcal{P}_J(s,t) \to 
\frac{\mathcal{P}_G}{(2\pi)^{1/2}(as)^{1/4}}e^{2\sqrt{as} - as}  \ll
\mathcal{P}_G .
\label{front}
\end{equation}
Thus, there is  a front at $s \sim s^{*}(t) = 1/a $ moving toward large $s$ as $t \to \infty$. The PDF ahead of this front is depleted
with respect to the Gaussian distribution, whereas behind the front {\color{black}it is asymptotically approaching}
to  Gaussian. Similar behaviour will be realised for any solution
(\ref{solnPgen}) evolving from an initial PDF whose decay at large $s$ is faster than exponential.

Further we shall also consider the moments
$$M^{(p)} = \langle |a|^{2p} \rangle \;\; (p=1,2,3,..),$$
so that the spectrum $n_{\bf k}$ is represented by the first moment ($p=1$),
whereas  the higher moments
 contain information about fluctuations of the wave-action spectrum about its mean value. For example, for the standard deviation we have
\BE
\sigma = (\langle |a|^4 \rangle - \langle |a|^2
\rangle^2)^{1/2} = (M^{(2)} - (M^{(1)})^2)^{1/2}.
\label{NazarenkoXI}
\EE
 If the wave amplitudes are deterministic, then $M^{(p)} = (M^{(1)})^p$ and $\sigma=0$.  For the
opposite extreme of large fluctuations we would have $M^{(p)} \gg (M^{(1)})^p$,
which means that the typical realization is sparse in the k-space and
is characterized by few intermittent peaks.
%

For Gaussian fields $ M^{(p)}=p!\ (M^{(1)})^p$.
%
To study the deviations from Gaussianity, it is useful to look at the relative cumulants 
%
%
\begin{equation}
\label{relCum}
F^{(p)} =\frac{ M^{(p)} - p! \, (M^{(1)})^p} {p! \, (M^{(1)})^p}, \qquad
 p=1,2,3,... 
 \end{equation}
By definition, $F^{(1)}$ is always zero. For $p=2$, this expression
measures the flatness of the distribution of Fourier amplitudes at
each $k$.  It determines the r.m.s. of the fluctuations of the wave-action $\sigma^2 = M^{(2)} -
(M^{(1)})^2  =  (M^{(1)})^2 (2 F^{(2)} +1). $
%
%

The general solution for the relative cumulants can be obtained
either from the general solution for the PDF, \eqref{solnPgen}, \eqref{solnPst1},
or by direct 
derivation of these quantities using \textcolor{black}{the weak WT turbulence}, as it was done in \cite{lvov_2004}.
The result is
\BE F^{(p)}(t) = e^{-p\theta} \sum^p_{j=2} {\theta^{p-j} p!  \over j!
(p-j)!}  F^{(j)}(t=0),
\label{fsoln}
\EE
where $\theta = \int_0^t {\eta \over n} dt'$ is an effective
time variable. One can see that expression \eqref{fsoln} decays
as $t \to \infty$ for any fixed $p$, provided that $\theta \to \infty$.
This condition for approaching to Gaussianity is equivalent to the respective condition
for the PDF solution, \eqref{cond}.

 It was noted in \cite{lvov_2004} that, even if the  $F^{(p)}$  eventually
decay to zero at each {\em fixed} $p$, their initial values propagate in $p$
without decay toward the larger values of $p$. 
This effect is interconnected with the propagation of
the PDF front toward larger $s$, which we discussed before.

Below, we are going to report on the numerical simulations of the GPE and the WKE, in which the described above predictions
about the PDF and the relative cumulants will be put to test alongside the comparisons of the wave spectra.

\section{Numerical methods} \label{sec:numerics}

\subsection{Gross-Pitaevskii equation}
\label{GPE_n}
The GPE (\ref{GPE}) is a partial differential equation with a cubic nonlinearity,
which \textcolor{black}{in the present work is considered in a 3D physical space and} integrated with the pseudo-spectral code FROST.
\textcolor{black}{The
complex field $\psi(\bm{r},t)$ is taken to be triply-periodic in the physical space with period $L$ and} represented on the grid of $N_p\times N_p \times N_p$ points. We evaluate the linear terms
in Fourier space and the nonlinear term in the physical space, which we then transform to Fourier space.
{\color{black}The classical $2/3$ rule is used for dealiasing \cite{gottlieb1977numerical}.}
The parallelization of the code uses the interface of
MPI communication and standard Fast Fourier Transforms (FFT) of the
FFTW library{\color{black}\cite{frigo2005design}.}
The library distributes the computation on various processors in
dividing the domain into slices.
For very large resolutions, a
partially  hybrid scheme is implemented, in addition with the OpenMP shared memory library. This
scheme allows to use more processors at fixed resolutions.
One can find more details about the FROST code in \cite{KrstulovicHDR}.

For integration on GPE with respect to time variable, we employ the exponential time differencing Runge--Kutta scheme of fourth order (ETD4RK) (see \cite{cox2002exponential}). In contrast to the classical Runge--Kutta scheme it provides  good  precision of approximation for both dissipative and dispersive partial differential equations (e.g., GPE as the latter), especially at the smallest scales of the Fourier space.

In the Fourier space we compute {\color{black}the spherically-integrated} wave-action spectra ${\color{black}n^{\text{rad}}}(k,t)$ defined by equation (\ref{eq:nk1D}) using its 
discrete space version:
\begin{equation}
{\color{black}n^{\text{rad}}}(k,t)=\frac{1}{D_k} \sum_{{\bf k}\in \Gamma_k } |\hat{\psi}({\bf k},t)|^2 \,.
\label{eq:GPE1D}    
\end{equation}
Here, $\Gamma_k$ is the spherical shell around $|{\bf k}|=k$ with the thickness of $D_k$, where $D_k$ equal to the mesh size $2\pi/L$ (spacing between the grid points in the Fourier space). 
\textcolor{black}{Such spherically-integrated spectrum represents the wave-action density in the 1D space of $k=|{\bf k}|$: it  is 
relevant to the situations when turbulence is (at least approximately) statistically isotropic.}

We run the code with resolutions $N_p=128$, $256$ and $512$ and fix the maximum $|{\bf k}|$ to be $k_{\text{cutoff}}$, so that the sizes of
\textcolor{black}{periodic}
box are $L=2\pi$, $4\pi$ and $8\pi$, respectively.
For each spatial resolution, an appropriate constant time step is set, such that the relative variation of the global invariants is less than $10^{-4}$.

A comment is due on the procedure of averaging in the GPE simulations.
Strictly speaking, for computing the ensemble averaged spectrum and the PDF of $J_{\bf k}$ generated by the GPE, one should perform a large number of numerical simulations starting from independent random initial fields. Performing such a task is unrealistic because it would require computational resources which are far beyond the ones presently available.
Instead of this, thanks to the \textcolor{black}{approximate } statistical isotropy, we get the statistical averaging via using all wavevectors contained in a given spherical shell in the spectral space. \textcolor{black}{Note that the statistical isotropy is present in the initial conditions, and it is approximately preserved throughout the simulations at the small scales (where the influence of the large-scale anisotropy due to the 3D periodic cube is minimal).}

\subsection{Wave-kinetic equation}

\label{KE_n}

The WKE is solved numerically by adapting the method developed in \cite{SemGreMedNaz} and by applying a second order Runge--Kutta scheme for time marching. The integrals of the collision term are computed using decomposition of the domain of integration into bounded subdomains, inside of which the integrands are highly-smooth functions. Each subdomain is mapped to the reference square, where we construct special grids. Coordinates of nodes of these grids are zeroes of Chebyshev polynomials. Then, we adapt and modify the  Clenshaw--Curtis quadrature to compute the integrals on such grids. {\color{black} This method is well suited for the collisional kernel of the GPE-based WKE, highly accurate and efficient} \cite{SemGreMedNaz}.

    For approximating the spectra on the specified range $[\omega_{min},\omega_{max}]\subset[0,\infty)$ we use a barycentric interpolation formula with Chebyshev nodes. In our study we set $\omega_{min}=0.1^2$, $\omega_{max}=86^2$.
The generalization of barycentric formula to the class of rational approximations is used as well. Such a method enables one to resolve possible singularities of spectra.
 \textcolor{black}{For computing collision term we use constant continuation of the function $n_{\omega}(t)$ to the segment $\omega\in[0, \omega_{min}]$ and assume that $n_{\omega}(t)=0$ for all $\omega>\omega_{max}$ and all $t$.} 
More details can be found in Appendix~\ref{Numerics}.  

Numerical solutions of the WKE are defined on a grid presenting a scale separation of almost six decades in $\omega$. The time step and the space grid are adjusted in order to ensure a good conservation of invariants. To address the issue of conservation, let us remind that the solution of the WKE is predicted to blow up at $\omega=0$ at a finite time $t^*$
\cite{semikoz1995kinetics, connaughton2004kinetic, lacaze2001dynamical, SemGreMedNaz}, which was estimated in our numerical tests as $t^*\approx 126.7$. For the times $t\in[0,100]$, which are far from $t^*$, we used the grid with $128$ nodes. The maximum relative deviations of the particle and energy densities from their initial values are $4.7\times 10^{-5}$ and $7.6\times 10^{-4}$, respectively. This non-conservation is mainly due to the leak of particles and energy through the boundary $\omega_{max}$. To compute the spectrum at the time $t=125.7$, which is rather close to $t^*$ (see Figure~\ref{figSpectraL}), we use a much finer grid with $1024$ nodes and expand the segment $[\omega_{min},\omega_{max}]$ to $[0.001, 100^2]$ in order to reduce the leak. In this case the biggest non-conservation $3.69\times 10^{-4}$ is observed for the particle density.
However, it is worth noting that this small non-conservation of the total particle number is accompanied with rather large approximation error, of the order 15\%, in the point-wise values of the wave action spectrum in the vicinity of $\omega_{min}$. This error is estimated by comparing the $1024$ and $512$ results.  This leads to conclusion that at the moment $t=125.7$, i.e. rather close to the blow-up $t^*$, our method still provides an adequate description. More details are available in Appendix~\ref{Numerics}.

Finally, to compare solutions of the WKE with those of the GPE, we notice that
\begin{equation}
\label{KE1D}
{\color{black}n^{\text{rad}}}(k,t) = 4\pi \omega n_{\omega}(t),
\end{equation}
where the dispersion relation $\omega = k^2$ was used.

\subsection{Probability distribution functions}
\label{PDF_n}

We shall concentrate our attention on the case when the
initial amplitudes are deterministic, so that the PDF
is given by the Green's function $\mathcal{P}_J(t,s)$.
We shall also consider the isotropic WT  and examine
the solutions at several different  $k=|{\bf k}|$.
Before computing $\mathcal{P}_J(t,s)$ for a given $k$ by formula \eqref{solnPst1},  it is convenient to normalize the PDF, so that the first moment of the new variable is equal to unity.
Here we introduce the normalized PDF $\tilde{\mathcal{P}}_J(t, \tilde s)=\langle s\rangle P_J(t, \tilde{s}\langle s\rangle )$ with the new stochastic variable $\tilde s = s/\langle s\rangle $, where $\langle s \rangle $   represents the ensemble average  of intensity $ J_{\bf k} $ at time $t$.
After this normalization the PDF corresponding to the Gaussian statistics becomes $\mathcal{P}_G(\tilde s)=\exp(-\tilde{s})$. 


To compute the  normalized PDF $\tilde{\mathcal{P}}(t,\tilde s)$, we first solve  the WKE numerically using the algorithm described in section~\ref{KE_n}, and then use the found spectrum to compute $\mathcal{P}_J(t,s)$ using \eqref{solnPst1}. To compute the temporal integral of $\gamma(t')$ in \eqref{nPDF}, we use a Gaussian quadrature formula with 9 nodes. We carefully check that the PDF is properly normalized, and the mean of $\tilde s$ is $1$ up to an error smaller than $3\times 10^{-4}$.


\section{Numerical results}\label{sec:NumResults}

The simulations of both GPE and WKE start with the following initial 
wave-action spectra of Gaussian shape,
\begin{equation}
\label{initialData}
{\color{black}n^{\text{rad}}}(k,0) = g_0\exp\biggl(\frac{-(k-k_s)^2}{\sigma^2}\biggr).
\end{equation}
The values of parameters $k_s$, $\sigma$ are chosen in such a way that the spectrum has enough space to spread to the left  before building up strong condensate at the zero mode. On the other hand, $k_s$ should not be too large that the direct cascade {\color{black}of energy \cite{nazarenko2011wave, ZLF}} can evolve for significantly long  time before the right front of spectrum 
touches $k_{\text{cutoff}}$ --- the right boundary of computational domain in the Fourier space. For these  reasons, we set $g_0=1$, $k_s=22$, $\sigma=2.5$ and $k_{\text{cutoff}}=43$ in the present work.

For WKE, we set $n_\omega(0)={\color{black}n^{\text{rad}}}(k,0)/4\pi\omega$ with $\omega=k^2$.

For GPE, we set $ \hat{\psi}_{\bf k} (0) = |\hat{\psi}_{\bf k}(0)| \exp(i\phi_{\bf k}(0))$ with  deterministic initial amplitudes $|\hat{\psi}_{\bf k}(0)|=\sqrt{\frac{{\color{black}n^{\text{rad}}}(k,0)}{4\pi k^2} (\frac{2\pi}{L})^3}$ and random initial phases $\phi_{\bf k}(0)$ uniformly distributed in $[0,2\pi)$ and statistically independent for each ${\bf k}$.
Such initial setting for GPE makes the RPA assumption satisfied at $t=0$ and also corresponds to the initial PDF of intensities $J_{\bf k}= |\hat{\psi}_{\bf k}|^2$ being a $\delta$-function. 

\subsection{GPE and WKE comparison for the short time evolution}

\subsubsection{Spectrum evolution}
\begin{figure}[htb]
\includegraphics{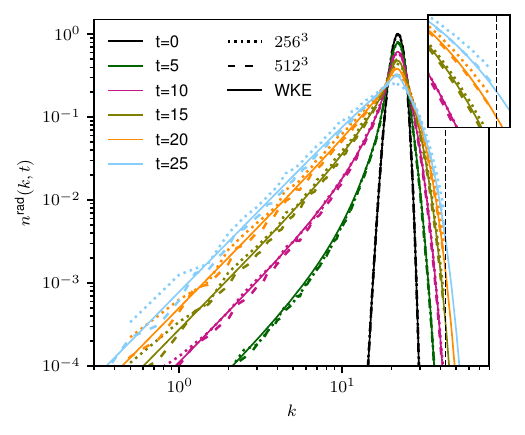}%
\caption{\label{figSpectraNs} Time evolution of ${\color{black}n^{\text{rad}}}(k,t)$ for short time. Results for GPE with $256^3$ and $512^3$ resolutions and WKE at  times $t=0,5,10,15,20,25$. The vertical dashed line marks the cut-off wave number for GPE numerics, $k_{\text{cutoff}}\approx 43$.}
\end{figure}

Let us compare the {\color{black} spherically-integrated} wave-action spectra, ${\color{black}n^{\text{rad}}}(k,t)$, from the GPE and the WKE simulations.  
The temporal evolution of ${\color{black}n^{\text{rad}}}(k,t)$ is plotted in Figure~\ref{figSpectraNs}.
{\color{black}
For the purposes of  the present work, we define  the nonlinear (kinetic) evolution timescale $T_{kin}$  as the time it takes for the maximum of spectrum to decrease to the half of its initial value, which in this study is 15.} 
First of all, we observe that the solutions of the  GPE and the WKE agree well until $t=25$.
Of course, the results of GPE obtained with $512^3$ resolution show better agreement with WKE than the results with $256^3$ resolution. At small $k$'s, the inverse cascade of the GPE solution obtained with $256^3$ resolution is initially slightly slower ($t<10$), and then slightly faster ($t>10$) than the inverse cascade of the WKE solution, whereas the spectra obtained from the GPE using $512^3$ resolution keep closely behind the ones of the WKE. In the ultraviolet region --- for large $k$'s --- the solutions of GPE with $256^3$ and $512^3$ resolutions start notably deviating from $t=20$, and the former deviates also from the solutions of the WKE. This can be seen from the zoomed plot presented at the right top corner of Figure~\ref{figSpectraNs}. The vertical dashed line in Figure~\ref{figSpectraNs} indicates the cut-off wave number  used in numerical method of solving the GPE, $k_{\text{cutoff}}\approx 43$. The cut-off wave number of the WKE is twice larger than the one of the GPE:  $k_{max}=\sqrt{\omega_{max}}=2k_{\text{cutoff}}\approx 86$.




Despite the remarkable agreement of the temporal evolution of the  GPE and the WKE wave-action spectra, some minor differences appears at large scales (small $k$'s), probably  due to the $k$-space discreteness. In order to give a more quantitative comparison between the solutions of the GPE and the WKE, we compute the wave-action and energy based centroids $K_N$ and $K_E$ \textcolor{black}{(the most particle- and energy-containing wave numbers respectively)} and their respective typical widths $\Delta_{K_N}$ and $\Delta_{K_E}$  in the interval $[0,k_{\text{cutoff}}]$ as follows
\begin{equation}
\label{centroids}
\begin{aligned}
&K_N(t)=\frac{1}{N_c(t)}\int_{0}^{k_{\text{cutoff}}}k{\color{black}n^{\text{rad}}}(k,t) \mathrm{d}k,\\
&\Delta_{K_N}(t)=\sqrt{\frac{1}{N_c(t)}\int_{0}^{k_{\text{cutoff}}}(k-K_N)^2{\color{black}n^{\text{rad}}}(k,t) \mathrm{d}k},\\
&K_E(t)=\frac{1}{H_c(t)}\int_{0}^{k_{\text{cutoff}}}k{\color{black}E^{\text{rad}}}(k,t) \mathrm{d}k,\\
&\Delta_{K_E}(t)=\sqrt{\frac{1}{H_c(t)}\int_{0}^{k_{\text{cutoff}}}(k-K_E)^2{\color{black}E^{\text{rad}}}(k,t) \mathrm{d}k}\,.
\end{aligned}
\end{equation} 
In the above definitions, ${\color{black}E^{\text{rad}}}(k,t)=k^2{\color{black}n^{\text{rad}}}(k,t)$ is the {\color{black} spherically-integrated    energy spectrum (full energy for the WKE and its linear-dynamics part for the GPE)}; $N_c(t)=\int_{0}^{k_{\text{cutoff}}}{\color{black}n^{\text{rad}}}(k,t) \mathrm{d}k$ and $H_c(t)=\int_{0}^{k_{\text{cutoff}}}{\color{black}E^{\text{rad}}}(k,t) \mathrm{d}k$.
For the GPE, $N_c(t)=N$ is constant, but $H_c(t)=H_2(t)$ varies around 1\% at $t=25$ (because GPE conserves the density of total energy $H$ including the nonlinear-dynamics energy). For the WKE, $N_c(t)$ and $H_c(t)$ vary, as well, since some waves cascade to modes with $k>k_{\text{cutoff}}$.

\begin{figure}[hbt]
\centering
\includegraphics{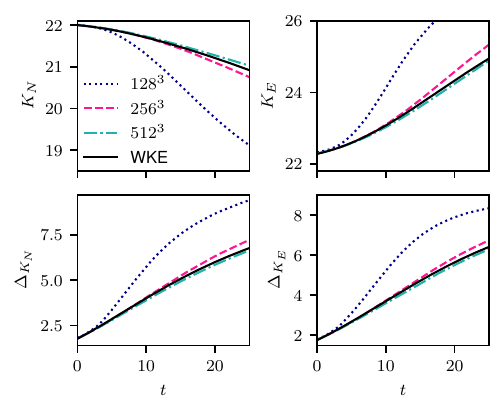}%
\caption{\label{figCentroids} 
Dynamics of the centroids of ${\color{black}n^{\text{rad}}}(k,t)$ and ${\color{black}E^{\text{rad}}}(k,t)$ and their respective typical widths  $\Delta_{K_N}$ and $\Delta_{K_E}$ for short time, $t\in[0,25]$.
Results for the GPE with $128^3$, $256^3$ and $512^3$ 
resolutions  and for the  WKE.}
\end{figure}
Figure~\ref{figCentroids} shows the dynamics of the centroids and the typical widths for $t\in[0,25]$. These quantities obtained by solving the WKE and the GPE with  resolutions  $128^3, 256^3$ and $512^3$. 
At very short times, $t<3$, the data obtained from the GPE simulations at all three resolutions and from the WKE demonstrate good agreement. The deviations between the solutions of the GPE with resolutions $256^3$ and $512^3$ remain rather small until much later times, $ t \sim 25$. Since a good accuracy is obtained by solving the GPE with $512^3$ resolution, in what follows all the GPE results are given with this resolution.

\subsubsection{Verifying the WT assumptions for the GP settings}
Here, the assumptions made in section \ref{sec:theory} for the four-wave weak WT will be discussed in the framework of the  setup chosen for the GPE simulations.  First of all, the condensate fraction $C_0$ is less than $5\times 10^{-7}$ until $t=25$, and the wave number corresponding to the healing length is $k_{\xi}=2.21$. This means that for almost all wave numbers ($k>k_{\xi}$) the influence of condensation is negligible and the nonlinearity remains small.

\begin{figure}[htb]
\includegraphics{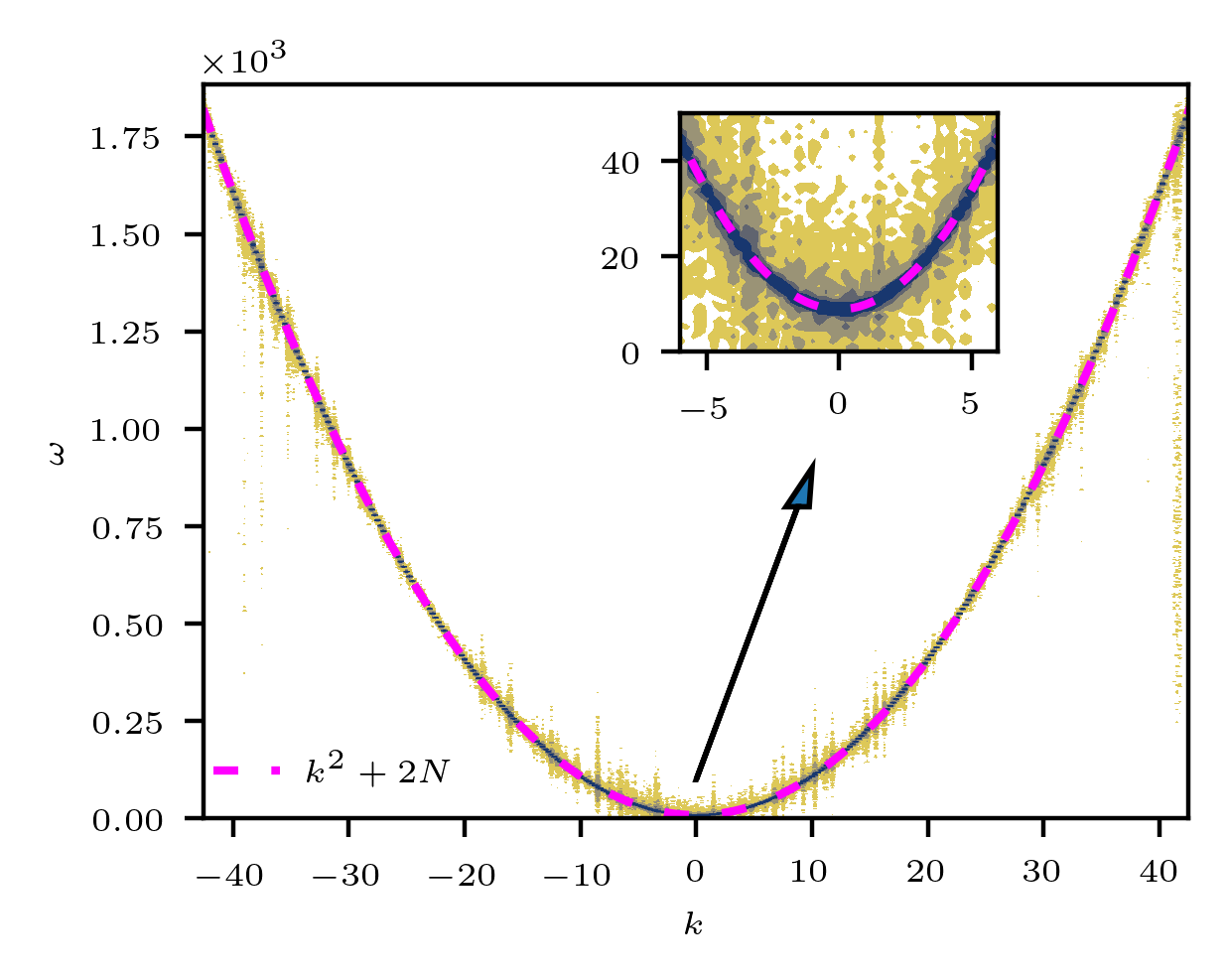}%
\caption{\label{figSpaTim} Spatio-temporal spectral density of $\psi(\bm{r}, t)$ over the time interval $[12,18]$.}
\end{figure}

\begin{figure}[htb]
\includegraphics{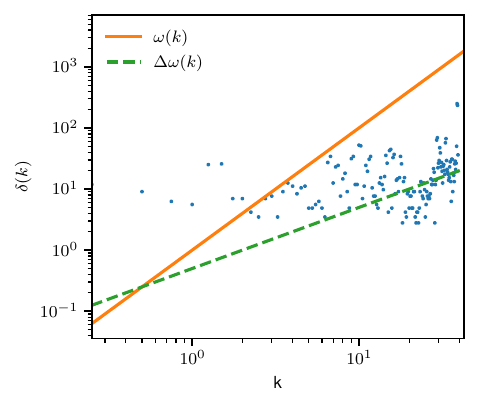}%
    \caption{\label{figdeltak} Frequency broadening $\delta(k)$ (blue points)  obtained from the spatial-temporal spectral density over the time interval $[12,18]$. }
\end{figure}

We compute the spatio-temporal  spectral density  of the wave function $\psi(\bm{r}, t)$ by performing Fourier transform of  $\hat{\psi}({\bf k}, t)$ with respect to the time variable  over a finite window of the size $T_w$. 
Thanks to the spatial isotropy of $\hat{\psi}({\bf k}, t)$, we choose $k_x=k_y=0$ and perform the Fourier transform in time for
each $k_z$. {\color{black}The spatio-temporal spectral density is nothing else but the surface of $|\hat{\psi}(k_x, \omega)|^2$. }

Figure~\ref{figSpaTim} displays the normalized spatio-temporal spectral density of $\psi(\bm{r}, t)$ over the time interval $t\in[12,18]${\color{black}, which is around $T_{kin}$}.
The normalization is performed along each line $k$=const in $k$-$\omega$ plane separately. It consists in division of the spectral density function at the given fixed $k$ by the value of its integral with respect to the variable $\omega$ taken over the interval $[0,\infty)$.
The figure shows that the majority of the spectrum is concentrated close to the  frequencies which 
satisfy the dispersion relation $\tilde{\omega}(k)=\omega(k)+2N$ for the corresponding wave numbers, see (\ref{eq:ST}).
It should be noted that  $\omega(k)=k^2$ is the 
linear-wave dispersion relation, and $2N$ is the shift induced by the nonlinearity (this shift is shown in the small plot). 

Moreover, because of the  nonlinearity, a broadening of the frequency can be observed
around $\tilde{\omega}(k)$: sufficiently narrow broadening implies weak-wave regime. 
One can measure the nonlinear frequency broadening $\delta(k)$ directly from the spatial-temporal  spectral density.
Here, we define $\delta(k)$ for each fixed $k$ in such a way that the integration of the spectral density over the interval of the width $\delta(k)$ centered at the $\omega$-peak gives the value $0.99$. To capture such information,
the length of the time window $T_w$ of FFT should be {\color{black}larger} than both linear and nonlinear time scales, which are $\frac{2\pi}{\omega(k)}$ and $\frac{2\pi}{\delta(k)}$ respectively. Meanwhile, $T_w$ should not be too large so that the spectrum does not 
vary much over $T_w$ to provide a good accuracy. 

Figure~\ref{figdeltak} presents the frequency  broadening $\delta(k)$ obtained from Figure~\ref{figSpaTim}. We choose the time
window $t\in[12,18]$ around $t=15$, when {\color{black}the} GPE gives good agreement with the WKE. The \textcolor{black}{weak WT theory}, on  one hand, 
requires the nonlinear time scales to be {\color{black} much greater than the linear ones (this amounts to the constraint  $\delta(k)<\omega(k)$)}. On the other hand, for solutions of the GPE in the discrete  Fourier space to be in the continuous $k$-space regime assumed by the \textcolor{black}{weak WT theory},
$\delta(k)$ should be greater than the frequency distance between the adjacent wave modes, $\Delta\omega(k) = 2k\Delta k$. This condition is needed to excite the nonlinear resonant and quasi-resonant interactions among waves \cite{nazarenko2011wave}.
\textcolor{black}{(For more discussion about the role of the quasi-resonant interactions see also \cite{PhysRevE.63.046306,zakharova2005mesoscopic,LVOV200624,dyachenko2003decay, annenkov2006role,PhysRevE.82.056322}.)} 
One can see that there is a significant $k$-range for which most of the points ($k, \delta_k$)
lay in the domain bounded by $\omega(k)$ and $\Delta\omega(k)$. 

It is interesting to see that $\delta(k)$ is greater than $\omega(k)$ at small $k$'s, 
which implies strong nonlinearity in the largest scales. However, Figure~\ref{figSpectraNs} shows that the wave-action spectrum is small in this range. It reminds us that the ``weak wave turbulence'' assumption exactly means weakly nonlinear waves rather than weak waves: even if a particular mode is very weak, the linear term at its wave number can be overpowered by the nonlinear term because the former is proportional to the frequency (which is small at small $k$'s) and the latter is enhanced by contributions from the other modes in the system.

One can also observe in Figure~\ref{figdeltak} that $\delta(k)$ sinks around 
$k=22$ (where the initial waves have maximum amplitudes) and some of the points fall even below the $\Delta\omega(k)$ line. 
This is because
the initial waves generate continuous 
cascades toward both low and large wave numbers.
When $\delta(k)$ falls below the $\Delta\omega(k)$ line, the discreteness of the $k$-space {\color{black}becomes significant} and deviations from the \textcolor{black}{weak WT theory} should be expected.

Thus we can see  in Figure~\ref{figdeltak} that the range of wave numbers where the \textcolor{black}{weak WT theory} assumptions are satisfied at $t\sim 15$ is, approximately, $2 \lesssim k \lesssim 20$.
This range becomes narrower for larger times.

\subsection{GPE and WKE  comparison for the long time evolution}

\begin{figure}[bt]
\includegraphics{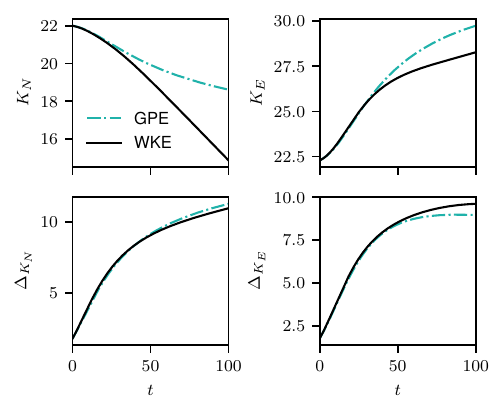}%
\caption{\label{figCentroidsL}Dynamics of the centroids of ${\color{black}n^{\text{rad}}}(k,t)$ and ${\color{black}E^{\text{rad}}}(k,t)$ and the corresponding typical widths.
Results for the $512^3$ GPE and the WKE for $t\in[0,100]$.}
\end{figure}

\textcolor{black}{Having observed a very good quantitative agreement between the numetical solutions of the GPE and the WKE at the times up to about two kinetic times, we would now like to prolong the computations to longer times to see how such an agreement gradually degrades, and if a qualitative similarity survives at the late stages}. In Figure~\ref{figCentroidsL} centroids and related typical widths are plotted up to $t=100$.
The centroids of wave-action spectra start to deviate a little earlier than the ones of energy spectra. At late times, the former exhibits larger deviations. The typical widths $\Delta_{K_N}$ and $\Delta_{K_E}$ 
show even better correspondence --- the curves obtained by the GPE and the WKE go very close until $t\approx 50$.
\begin{figure}[bt]
\centering
\includegraphics{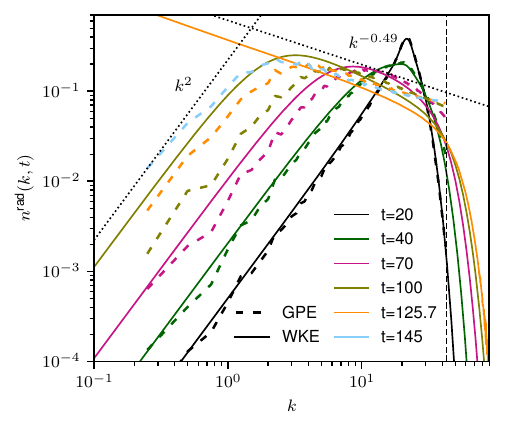}%
\caption{ \label{figSpectraL}Long term evolution of ${\color{black}n^{\text{rad}}}(k,t)$. Results for the $512^3$ GPE and the WKE at $t=20,40,70,100,125.7$ respectively and for the GPE at $145$. }
\end{figure}

The comparison of the spectra obtained by solving the GPE and the WKE  until $t=100$ is given in Figure~\ref{figSpectraL}. The disagreement between the GPE and the WKE takes place at both low values and high values of $k$ {\color{black}starting} around $t=40$.  Overall, the GPE is faster than the WKE for the direct cascade, but slower for the inverse cascade. This trend can be also observed in Figure~\ref{figCentroidsL} --- the evolution of $K_N$ of the GPE to small $k$'s is slower than the evolution of $K_N$ of the WKE, whereas the evolution of $K_E$ to large $k$ is faster.

\textcolor{black}{The observed deviations between the long-term GPE and WKE spectra can be attributed to the following major factors: (i) failure of the conditions of the WKE applicability due to  increased nonlinearity or/and  increasing importance of the finite-box effects, and (ii)   numerical effects due to differences of computational ranges and the discretization steps in the GPE and the WKE codes. Below, we will discuss these effects while trying to separate explicitly which one of them is the "failure of the weak WT theory" and which is "numerical".  }

\textcolor{black}{
First of all, we note that the deviations are totally expected for the spectrum propagating into the region of small wave numbers because of the failure of the weak WT theory in this range. Indeed,  according to Figure~\ref{figdeltak}, even at $t\sim 15$ the nonlinear frequency broadening $\delta(k)$ is greater than the linear-wave frequency for $k\le 2$ so the WKE description is not applicable there. Since the spectrum at low $k$ is growing, this  boundary of WKE applicability moves to higher $k$'s at later times. }

An interesting and important fact is that, even though the GPE is slower than the WKE for small values of $k$, both the GPE and the WKE
obey the global thermodynamic equilibrium scaling   $k^2$ (the left black {\color{black}dotted} line), but with different characteristic evolution times. 
Then considering the long-time evolution, it is important to keep in mind that the solution of the WKE is predicted to blow up at $k=0$ at a finite time $t^*$
\cite{semikoz1995kinetics, connaughton2004kinetic, lacaze2001dynamical, SemGreMedNaz}, so that the WKE cannot be computed beyond this time without a modification of the model taking into account the zero-mode evolution. For the initial condition  considered in the present paper, we found numerically that $t^*\approx 126.7$.
Close to $t^*$, the evolution appears to be self-similar, with a transient power law forming in the inverse cascade {\color{black}$n^{\text{rad}}(k) \propto k^{-x^*}$} with exponent $x^*$ which has a
different value than the one in the inverse-cascade KZ solution, namely $x^*$ differs from $1/3$.
The most careful recent study of the self-similar formulation of the WKE \cite{SemGreMedNaz} gives two values of $x^*$ for the most accurate solutions,  $0.44$  and $0.48$. Further, a similar behaviour was observed in $512^3$ simulations of the 3D GPE with a slightly different value, 
$x^* \approx 0.52$ \cite{vish}. In Figure~\ref{figSpectraL}  we see that the late WKE spectra
exhibit law  $k^{-0.49}$ (indicated 
by a black {\color{black}dotted} line)  whereas the GPE scaling is slightly less steep, probably due to the spectrum pileup near $k_{\text{cutoff}}$.


Let us now concentrate on the differences that the GPE and the WKE exhibit for the inverse and the forward cascade ranges \textcolor{black}{that may be attributed to the differences in the numerical setups for the GPE and WKE simulations}. It is easy to see that for small wave numbers (i.g. $k<10$), the curves obtained by the GPE show visible fluctuations starting from $t=40$, while the results generated by the WKE keep good smoothness all the time. 
This behaviour is to be expected considering the fact that the WKE deals with ensemble averaged wave distributions and with a space of continuous wave number (allowing for using the grids in $k$ variable, which are finer at low $k$'s), whereas the GPE spectra are not ensemble (or time) averaged: the only averaging in this case is over spherical shells in the Fourier space, which contain less and less modes as one moves to smaller $k$'s.

\textcolor{black}{Note that, even though the WKE grid can be refined close to $\omega=0$, the WKE does not include the evolution of a condensate mode, unlike the GPE.} Indeed, while the GPE solution  evolves, the condensate fraction $C_0$ may start to grow. However, in the current simulations its final value is less than $5\times10^{-5}$ at $t=145$, so the influence of condensate in the GPE simulation still can be neglected. The fact that the WKE predict condensation starting at $t^* \approx 126.7$ whereas no condensate growth is seen in the GPE simulation even
at $t=145$ could be qualitatively understood if we recall  the condensation criterion for the weakly interacting 3D GPE: $E/N < k_{max}^2/3$ \cite{Connaughton2005}.  It turns out that for our initial condition this criterion is well satisfied for the WKE but only marginally so for the GPE because
the former has a twice bigger cutoff wave number than the {\color{black}latter}. In a separate simulation we have verified that condensation does indeed occur for the GPE too if the initial spectrum is shifted to smaller wave numbers (leading to smaller $E/N$).
%

On the other side, the {\color{black}high wave number} cut-off of the GPE results in \textcolor{black}{numerical effect of} accumulating of waves near $k_{\text{cutoff}}$,  which affects the overall dynamics {\color{black}of} the system.
This is the well-known ``bottleneck" phenomenon
typical for many turbulent systems which could roughly be described as the energy flux stagnation at scales slightly larger than the scales where the spectrum is depleted either by a dissipation (especially  if the latter is of the  hyperviscous type) or by  a numerical cutoff (like in our case) \cite{Falkovich,borue1995self,Cichowlas,Dispersive}.
It is difficult to break this restriction in numerical simulations since it is induced by finite computational domain and finite space resolutions. However, the WKE can provide solutions for large values of $k$ with acceptable requirement of computational resources. 

\textcolor{black}{Finally, let us comment on the effect of discreteness of the wave number space in the GPE simulations due to the finite size of the periodic box $L$ -- an effect which is not present in the WKE formulation whose frequency space is continuous. Naively one could guess that this effect is most important at the low-$k$ range. However, 
as seen in Figure~\ref{figdeltak} for $t\sim 15$, the $k$-space discreteness becomes a problem at the high $k$'s and not the small ones. This is because at low $k$'s the nonlinear frequency broadening remains greater than the spectral distance between the adjacent linear eigenmodes
(which is $2\pi/L = 1/4$ in the $512^3$ runs). 
This picture persists for $t > 15$ and, moreover, the discrete effects become less important  at high $k$'s too because of the bottleneck spectrum accumulation in this region leading to increasing of the frequency broadening. To summarise, we believe that the finite-size effects are less important for understanding of the deviations between the GPE and the WKE results compared to the other effects we have mentioned above. }

\subsection{Low and high order statistics}

\subsubsection{Comparison of the PDFs and the  low order cumulants (GPE and WKE)}

\begin{figure*}[htb]
  \centering
  \begin{subfigure}[b]{0.33\textwidth}
  \centering\includegraphics[width=\textwidth]{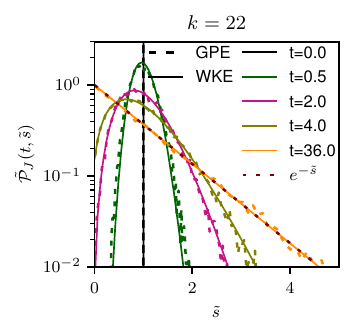}
  \end{subfigure}\hfill
   \begin{subfigure}[b]{0.33\textwidth}
  \centering\includegraphics[width=\textwidth]{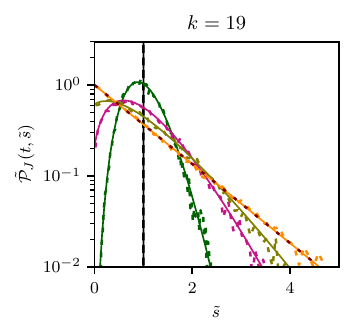}
  \end{subfigure}\hfill
  \begin{subfigure}[b]{0.33\textwidth}
  \centering\includegraphics[width=\textwidth]{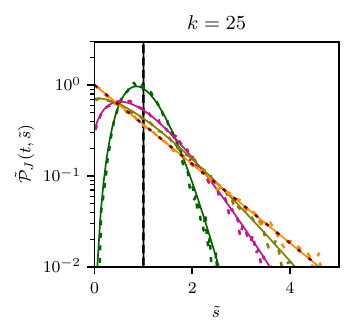}
  \end{subfigure}
  \caption{\label{figPDFcomp} Time evolution of the normalized probability density functions $\tilde{\mathcal{P}}_J(t,\tilde{s})$.
  Results for $512^3$ GPE and WKE for various $k$.}
\end{figure*}

\begin{figure*}[htb]
  \begin{subfigure}[b]{0.33\textwidth}
  \centering\includegraphics[width=\textwidth]{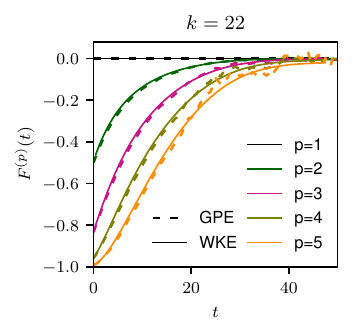}
  \end{subfigure}\hfill
   \begin{subfigure}[b]{0.33\textwidth}
  \centering\includegraphics[width=\textwidth]{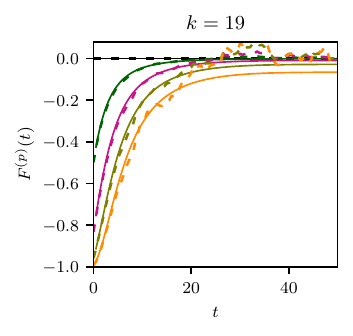}
  \end{subfigure}\hfill
  \begin{subfigure}[b]{0.33\textwidth}
  \centering\includegraphics[width=\textwidth]{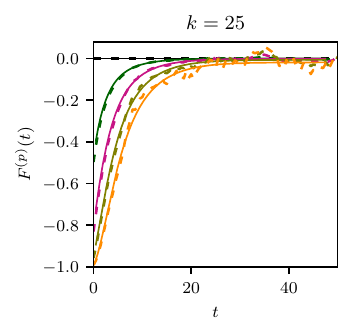}
  \end{subfigure}
  \caption{\label{figCumcomp} Time evolution of the relative cumulants $F^{(p)}(t)$.
  Results for $512^3$ GPE and WKE  for various $k$. }
  \end{figure*}

{\color{black}Let us now consider} the PDF of the transient states starting from the deterministic initial amplitudes and evolving towards an exponential PDF corresponding to a Gaussian wave-field statistics. We shall use here the previous setup \eqref{initialData} and analyze the evolution of the normalized PDF and the relative cumulants at the point $k=k_s=22$ --- the maximum point of the initial spectrum, and the points $k=19$ and $k=25$ symmetrically placed around $k=22$. Figures~\ref{figPDFcomp} and~\ref{figCumcomp} display the temporal evolution of
$\tilde{\mathcal{P}}_J(t,\tilde{s})$ and $F^{(p)}(t)$ respectively.
As expected, for all three modes the normalized PDFs for both the GPE and the WKE spectra evolve from the $\delta(\tilde{s}-1)$ at $t=0$ to the final exponential distribution $e^{-\tilde{s}}$. Overall, the results generated by the GPE agree well with the prediction derived from the WKE.
Also, from Figure~\ref{figPDFcomp} we see that the PDF at $k=22$ evolves towards Gaussianity slower than the PDFs at 
$k=19$ and $k=25$. This {\color{black}is consistent} with our prediction that the rate of convergence to 
Gaussianity is faster for those $k$'s where the initial spectrum $n_k(0)$ has smaller values.

Figure~\ref{figCumcomp} provides further validation of the statistical predictions  derived from the WKE
by plotting the relative cumulants of the first  to  fifth orders.  
All the cumulants finally tend to zero---the state corresponding to the Gaussian wave-field statistics (again, faster at $k=19$ and $k=25$ than at $k=22$).  For high orders, GPE cumulants start to deviate from those of WKE after reaching a certain time. As explained above, the discreteness of the GPE modes does not allow the degree of averaging sufficient for generating accurate PDF especially for large values of $\tilde{s}$. The accuracy of high order moments is also affected by such a discretization earlier in time than the accuracy of the low order ones. As a consequence, 
in Figure~\ref{figCumcomp} for $t<20$, the GPE gives good picture of $F^{(p)}(t)$ up to the fifth order at $k=22$, but at $k=19$ and $k=25$ good agreement with the WKE is preserved only up to $p=4$.

\subsubsection{High-order \textcolor{black}{statistics using the }WKE}


Studying high-order statistics for a time dependent problem requires to produce ensemble realizations in order to collect the statistics. For the GPE simulations presented in this work, such study is prohibiting. Instead, the \textcolor{black}{weak WT theory} provides a framework in which the high-order statistics can be studied. 

\textcolor{black}{To compute PDF, we first obtain the numerical solution of WKE, $n(t)=n_{\omega}(t)$. Then, we use the exact solution of evolution equation \eqref{main} for PDF, i.e. formulae \eqref{solnPgen}--\eqref{nPDF}, where the deterministic initial data  $J$ has Gaussian shape as a function of $k$, \eqref{initialData}. For computing the higher-order cumulants $F^{(p)}(t)$ we use formula \eqref{relCum}, where moments $M^{(p)}$ are the integrals of functions $\tilde{s}^p\tilde{\mathcal{P}}_J(\tilde{s},t)$. Thus, knowing the numerical solution of WKE in the framework of weak WT theory, one can restore the high-order statistics using exact formulae.}

\textcolor{black}{The goal of this section is to represent the transition between deterministic initial data with Gaussian $k$-distribution as a motion of fronts with respect $\tilde{s}$ and $p$ variables. Ahead of the fronts, PDFs and cumulants significantly deviate from the Gaussian statistics, whereas behind the fronts they are very close to the Gaussian statistics.} 

\textcolor{black}{To define such fronts, we set a threshold value of the relative deviation of the statistics from Gaussian equal to 1\% since this value becomes noticeable on the plots. Then, for each time moment $t$ we find the points $\tilde{s}$ and $p$, for which the relative PDF $\tilde{\mathcal{P}}(t,\tilde{s})/\mathcal{P}_G(\tilde{s})$ and cumulant $F^{(p)}(t)$ deviate from corresponding curves of Gaussian statistics by 1\%. In what follows, for these points we use  notations $\tilde{s}^*(t)$ and $p^*(t)$, respectively.}

\begin{figure}[h]
  \centering
  \begin{subfigure}[b]{0.5\textwidth}
  \centering\includegraphics[scale=0.55]{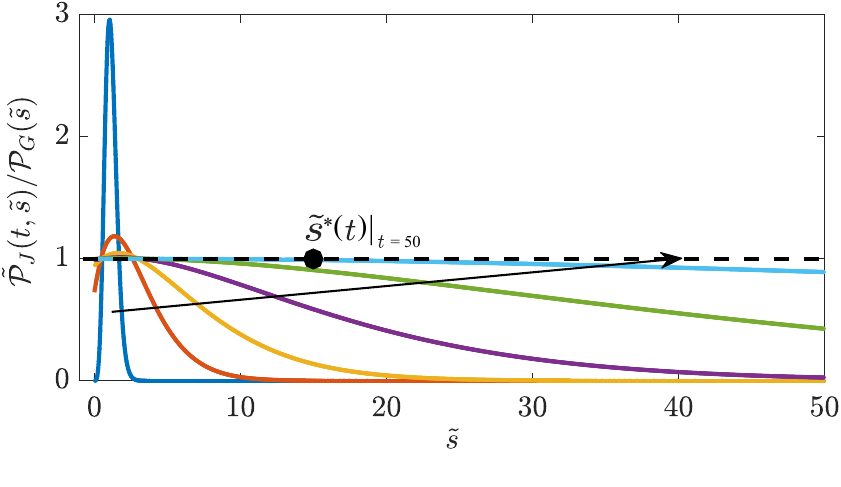}
  \caption{}\label{figPDF}
  \end{subfigure}\hfill
  \begin{subfigure}[b]{0.5\textwidth}
    \centering\includegraphics[scale=0.56]{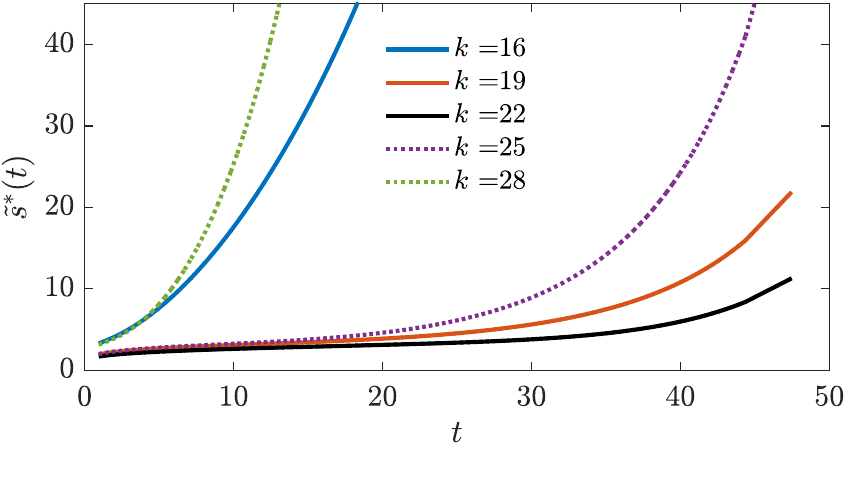}
    \caption{~}\label{figPDFfront}
    \end{subfigure}
\hfill
  \caption{\label{figPDFCumu} (a) Relative probability density functions computed using the numerical solutions of WKE  for different time moments and $k=22$. The arrows show time evolution and pass from the left to the right through the curves corresponding to $t=1$, $t=10.8$, $t=20.6$, $t=30.4$, $t=40.1$, $t=50$. Horizontal dashed line shows the Gaussian distribution. (b) Motion of the front $\tilde{s}^*(t)$ for different values of $k$. 
  }
 \end{figure}

Figure~\ref{figPDF} shows the relative PDFs computed at $k=22$. The black dot denotes the point $\tilde{s}=\tilde{s}^*(t)$ where the curve corresponding to $t=50$ \textcolor{black}{deviates} from the limiting dash line by 1\%, i.e. where the ratio of PDF \textcolor{black}{and exponential  PDF $\mathcal{P}_G(\tilde{s})$} takes the value $0.99$. The dynamics of the point $\tilde{s}^*(t)$ illustrates a front moving towards large $\tilde{s}$ as $t\to\infty$. Behind this front, PDF  \textcolor{black}{asymptotically approaches} to the exponential one, whereas ahead of this point it still deviates by more than $1\%$. Figure~\ref{figPDFfront} displays the dynamics of $\tilde{s}^*(t)$ for different values of $k$, showing its acceleration in the final phase, which depends on the wave number.

%

\begin{figure}[tb]
\centering
\begin{subfigure}[b]{0.5\textwidth}
  \centering\includegraphics[scale=0.55]{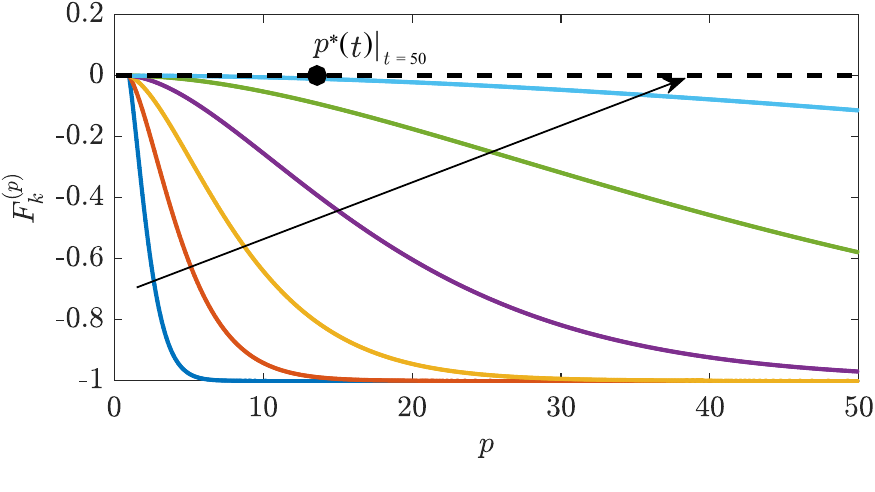}
  \caption{~}\label{figCumu}
  \end{subfigure}
\vfill
\begin{subfigure}[b]{0.5\textwidth}
\centering\includegraphics[scale=0.56]{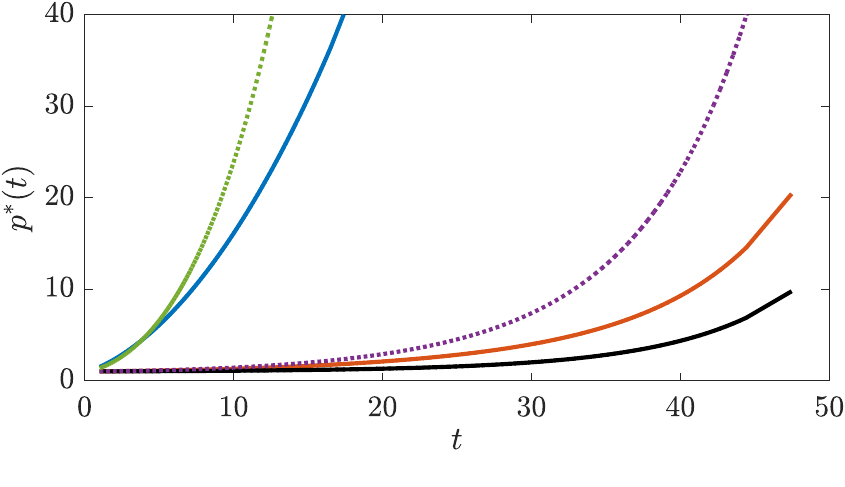}
\caption{~}\label{figCumufront}
\end{subfigure}
\caption{\label{figFront} (a) Cumulants as a function of their order $p$ (same times as in Figure~\ref{figPDFCumu}a). (b) Motion of the front $p^*(t)$ for different values of $k$ (as in Figure~\ref{figPDFCumu}b). Bottom black line corresponds to $k=k_s$. 
}
\end{figure}

\textcolor{black}{Let us now consider the behavior of the high-order relative cumulants.} In Figure~\ref{figCumu} the cumulants as functions of the order $p$ are shown at different times for $k=22$. Similarly, the point $p^*(t)$ corresponds to the value of $p$ for which at the time moment $t$ the cumulant $F_k^{(p)}$ takes the value $-0.01$. Figure~\ref{figCumufront} shows the propagation of front associated with the motion of the point $p^*(t)$.   

We should mention here an obvious similarity between the evolution of relative PDFs as  functions of $\tilde s$ and of the cumulants as  functions of $p$, when $t$ is large enough.

\section{Discussion and conclusions}\label{sec:conclu}

The main goal of the present paper is to test the weak Wave Turbulence theory  by comparing numerical simulations of the Gross-Pitaevskii equation (GPE) and the wave-kinetic equation (WKE). Several fundamental constraints have to be kept in mind during such a comparison. 

Firstly, the \textcolor{black}{weak WT theory} assumes that the wave must be weak enough, but not too weak. Namely, the waves have to be weak so that the linear dynamics is faster than the nonlinear transfers of energy between the wave modes, but also strong enough so that the nonlinear frequency broadening is larger than the frequency spacing between the nearby modes in the wave number space, so that the continuous $k$-space limit can be considered. We have checked these conditions in our numerical simulations. Figure~\ref{figdeltak} shows that the range of wave numbers, where these two \textcolor{black}{weak WT theory} assumptions are satisfied simultaneously at $t\sim 15$, is $2 \lesssim k \lesssim 20$, and we found that this range becomes narrower for larger times.
It is then rather remarkable that we found a nearly perfect agreement between the WKE and the GPE results up to $t\sim 30$, {\color{black}which is about two nonlinear (kinetic) evolution times.}
To characterise our comparisons, we have looked directly at the spectra, as well as at their global characteristics -- the energy and the particle centroids and typical widths. Another surprising fact is that even relatively low $256^3$ resolution of GPE simulations leads to a reasonably accurate  agreement between the WKE and the GPE results, even though, naturally, the accuracy is better for $512^3$
(and not at all good for $128^3$).

We have also looked numerically at the probability density functions (PDFs) of the wave intensities, both using GPE simulations (directly) and the \textcolor{black}{weak WT theory} ({\color{black}using} the analytical solution for the PDF obtained in \cite{Choi_2017}). Here as well, we have seen an excellent agreement for the early evolution up to $t \sim 30$ and, qualitatively, far beyond this time. In fact, in agreement with the theoretical 
predictions, the PDFs of the wave intensities  tend to  exponentially decreasing functions corresponding to Gaussian wave fields. At $t \sim 30$, the PDFs have practically attained the final
exponential shape, and the only difference between the \textcolor{black}{weak WT theory} and the GPE results is that the PDFs obtained for the GPE have   noisy tails due to insufficient averaging. It is important to emphasize that, in agreement with the theoretical predictions \cite{choi_2005b,Choi_2017,nazarenko2011wave}, the evolution toward Gaussianity occurs at the kinetic timescale (in our case $t\sim 15$) i.e. at the same characteristic time as the one for the spectrum evolution, rather than a much shorter timescale.
\textcolor{black}{On the other hand},  evolution of the wave intensity PDF from deterministic to exponential in simulations of 2D GPE was previously observed in \cite{Chibbaro}, where it was reported that PDF evolution time is much faster than the one for the spectrum, which \textcolor{black}{at the first sight} seems at odds with our findings. \textcolor{black}{This ``contradiction"} seems to originate from the difference in the definition of the characteristic evolution time for the spectrum: in the present paper we define it as a time needed for the spectrum to experience order-one changes (specifically, for its maximum to become twice smaller) whereas a much longer time needed for the spectrum to approach to thermodynamic equilibrium is discussed  in \cite{Chibbaro}. \textcolor{black}{With this remark in mind, we see that our results do not contradict the ones of \cite{Chibbaro}: the PDF core to the exponential shape at the kinetic timescale (even though the PDF tails and the cumulants evolve slower, see Figures \ref{figCumcomp}, \ref{figPDFCumu} and \ref{figFront}) whereas the spectrum, while evolving initially at the kinetic timescale (by definition) keeps evolving for much longer up to the blowup time. Note also, that the statement about the PDF evolution time has to be adjusted in the case of sharply peaked PDFs. From \eqref{main}, for the PDF core ($s\sim n$) we have the following estimate for the characteristic PDF evolution time, $T_{pdf} \sim T_{kin} (n/\delta s)^2$ where $\delta s$ is the width of the PDF peak.
In particular, our delta-shaped initial PDF means  that formally the initial PDF evolution time is zero.  }
We would also like to mention a previous study of a model 2D three-wave system \cite{Tanaka} where a comparison was made for the PDF and cumulants obtained from a DNS and the numerics of the  \textcolor{black}{weak WT } closure equations for these statistical objects \textcolor{black}{up to $t\sim 0.3 T_{kin}$} (no analytical solution for the PDF was available at that time yet). They arrived at favourable conclusions for validity of the \textcolor{black}{weak WT theory} for the considered three-wave system.

Secondly, \textcolor{black}{weak WT theory} assumes a ``propagation of chaos" property: the random phase and amplitude (RPA) statistics of the initial data should survive through the nonlinear (kinetic) evolution time \cite{choi_2004,choi_2005,nazarenko2011wave}. Naively, one could 
think that the RPA statistics could be tested numerically by directly accessing the joint statistics of the Fourier modes.
However, it is clear that RPA cannot propagate in its pure form for the entire set of the wave modes. Instead, it should survive only in a sense of distributions, e.g. the statistical moments and the reduced PDFs restricted to smaller number of waves \cite{choi_2005b,nazarenko2011wave}. Numerical studies of these issues would be a good subject for future research. In the present work, however, we see an indirect evidence for the propagation of chaos in the fact that WKE provides a good description of the wave spectrum for at least two nonlinear kinetic times.

It is interesting that a qualitative agreement of the GPE and WKE results is also observed to the long-time evolution, up to $t\sim 100$ and even beyond (but obviously for $t<t^*$, where $t^*$ is the finite time moment when the solution of WKE blows up at zero mode). In particular, in both simulations a self-similar inverse cascade with two power-law scalings (with exponents $\sim 2$ and $\sim -0.49$) is observed for $t$ close to $t^*$.
However, 
significant differences arise at the lowest and the highers scales for $t \gtrsim 40$. In addition to the shrinking of the \textcolor{black}{weak WT theory} applicability range, which we mentioned before, the factors causing the deviations include the differences in the minimal and the maximal scales used for the WKE and GPE. Generally, one can afford a much greater range of wave numbers in WKE than in GPE. In our WKE simulations we had the minimal wave number at about $0.1$ and the maximal wavenumber -- at $86$, whereas in $512^3$ GPE these boundaries where at $0.25$ and $43$ respectively. Having lower maximum wave number leads to a visible accumulation of spectrum in the highest wave numbers, whereas in the WKE results the forward cascade spreads freely to much higher $k$'s. It is then surprising that the energy centroid for the WKE appears to move to the right slower than the one for the GPE. It is equally surprising to see the inverse cascade moving toward low $k$'s faster for \textcolor{black}{the WKE than for the} GPE. Indeed, this result is at odds with a common view that the nonlinearity becomes stronger during the inverse cascade process, so the GPE system should switch from the kinetic to a dynamic timescale which is shorter. 
At present, we do not have an explanation for this behaviour.

\textcolor{black}{Once again, we would like to reiterate our
view that the late-time deviations between the GPE and the WKE evolution are due to a combination of two equally important factors: breakdown of WKE validity conditions and the existence of the high-frequency cut-offs which were different for the GPE and the WKE simulations.}

{\color{black} 
The  present work was restricted to the situations when the condensation fraction remains negligible.
A direct consequence of the condensation is that  the system  goes into a three-wave regime -- namely the acoustic wave turbulence. The corresponding WKE for waves on background of a strong background condensate was derived  in \cite{dyachenko1992optical}. 
Numerical simulations of GPE  were  previously performed for such a regime in \cite{proment2009energy,proment2012sustained,fujimoto2015bogoliubov} aiming at comparing their results to the  stationary KZ solution of the respective three-wave WKE. However, in future it would be interesting to study evolving WT in this regime and confront the GPE and the WKE results in the way similar to how it was done in the present paper aiming at testing validity of the WT approach in presence of a strong condensate component.}

\section{Acknowledgements}
This work is funded by the Simons Foundation Collaboration grant Wave Turbulence (Award ID 651471). Part of this work was granted access to the high-performance computing facilities under GENCI (Grand Equipement National de Calcul Intensif) A0102A12494 (IDRIS and CINES), the OPAL infrastructure from Université C\^{o}te d’Azur, supported by the French government, through the UCAJEDI Investments in the Future project managed by the National Research Agency (ANR) under reference number ANR-15-IDEX-01, and  the SIGAMM infrastructure (http://crimson.oca.eu) hosted by Observatoire de la C\^{o}te d'Azur and  supported by the Provence-Alpes C\^{o}te d’Azur region. 

\appendix

\section{Derivation of the angle-averaged kinetic equation}
\label{KEder}

In the case when the wave fields are statistically
isotropic, the averaging of WKE  \eqref{KE0} can be done by integrating it over the unit sphere $S_1$ and performing the internal angle integrations. This will change the variables ${\bf k}$ and ${\bf k}_j$, $j=1,2,3$, of \eqref{KE0} to  variables $k=|{\bf k}|$ and $k_j=|{\bf k}_j|$ respectively. In what follows we also use the representations $d {\bf k} = k^2dkd\Omega$ and $d {\bf k}_j = k_j^2dk_jd\Omega_j$, where $d\Omega$ and $d\Omega_j$ are the area elements on the unit spheres associated with the vectors ${\bf k}$ and ${\bf k}_j$, respectively. Isotropy of the spectrum means that it depends only on the modulus of the vector ${\bf k}$, i.e. $n_{\bf k} = n_k$.

Denoting $n_j=n_{{\bf k}_j}$, $\omega=\omega_{{\bf k}}$ and $\omega_j=\omega_{{\bf k}_j}$ one can write the averaged WKE as follows
\begin{equation}
\label{WKEav}
\begin{aligned}
&\frac{dn_k}{dt} = \frac{1}{4\pi}\int_{S_1}\frac{dn_{\bf k}}{dt}d\Omega = \\ 
&=\int \mathcal{P}\delta_{1\omega}^{23}n_kn_1n_2n_3 \bigg(\frac{1}{n_k}+\frac{1}{n_1}-\frac{1}{n_2}-\frac{1}{n_3}\bigg)d123,
\end{aligned}
\end{equation}
where
$$
\delta_{1\omega}^{23} = \delta(\omega+\omega_1 - \omega_2-\omega_3),~~d123=k_1^2k_2^2k_3^2dk_1dk_2dk_3,
$$
\begin{equation}
\label{Peq}
\mathcal{P} = \int \delta_{1{\bf k}}^{23} d\Omega d\Omega_1d\Omega_2d\Omega_3,
\end{equation}
where $\delta_{1{\bf k}}^{23} = \delta({\bf k}+{\bf k}_1-{\bf k}_2-{\bf k}_3)$.
To compute $\mathcal{P}$, we first recall that the Dirac $\delta$-function can be represented as the Fourier transform of unity:
\begin{equation}
\label{AI_Four}
\delta_{1{\bf k}}^{23}=\frac{1}{(2\pi)^3}\int\limits_{\mathbb{R}^3}\exp[-i{\bf r}\cdot({\bf k}+{\bf k}_1 - {\bf k}_2-{\bf k}_3)]d{\bf r},
\end{equation}
Therefore, substituting \eqref{AI_Four} into \eqref{Peq} and changing the order of integration one obtains
\begin{equation}
\label{deltaAver}
\mathcal{P}=\frac{1}{(2\pi)^3}\int\limits_{\mathbb{R}^3} P({\bf r})d{\bf r},
\end{equation}
where 
\begin{equation}
\label{Pexpr}
\!\! P({\bf r}) = \! \int \! \exp[-i{\bf r}\cdot({\bf k}+{\bf k}_1 - {\bf k}_2-{\bf k}_3)]d\Omega d\Omega_1 d\Omega_2 d\Omega_3.
\end{equation}

Let us pass to the angular coordinates of spherical surfaces, $(\varphi,\theta)$, where $0\leq\varphi\leq 2\pi$,  $0\leq\theta\leq \pi$ (and the same for $\varphi_j$, $\theta_j$):
$$
d\Omega=\sin\theta d\theta d\varphi,~~~d\Omega_j=\sin\theta_j d\theta_j d\varphi_j,~~~j=1,2,3.
$$

Defining $\theta_j$ to be to the angles between the vectors ${\bf k}_j$ and ${\bf r}$, we have
$$
{\bf r}\cdot {\bf k}_j = { r}{ k}_j\cos\theta_j,~~~j=1,2,3,
$$
where $r=|{\bf r}|$.
Similarly,
$$
{\bf r}\cdot {\bf k} = { r k}\cos\theta.
$$

Thus, one can write 
\begin{equation}
\label{AI_int1}
\begin{aligned}
&\int\limits e^{-i{\bf r}\cdot {\bf k}}d\Omega = \int\limits_{0}^{2\pi} d\varphi \int\limits_{-1}^{1} d \chi e^{-i r k\chi}=\\
& - \frac{2\pi}{i r k}\exp(-i r k\chi)\bigg|_{\chi=-1}^{1}=\frac{4\pi}{ r k}\sin({ r k}),
\end{aligned}
\end{equation}
where $\chi = \cos\theta$.

Using \eqref{AI_int1} and the similar results for the 
integrals of $e^{-i{\bf r}\cdot {\bf k}_j}$, $j=1,2,3$, one can write for the integral in \eqref{Pexpr}:
$$
P({\bf r}) = \frac{(4\pi)^4}{r^4}\frac{\sin(r k)}{k}\prod\limits_{i=1}^3\frac{\sin( r k_j)}{k_j}.
$$
Substituting this into \eqref{deltaAver}  using the representation $d{\bf r}= r^2d r\, d\Omega_{\bf r}$, and using {\it Mathematica} for integration,  we get the following expression,

\begin{equation*}
\begin{aligned}
&\mathcal{P} = \frac{32\pi}{kk_1k_2k_3}
\int d\Omega_{\bf r}\int\limits_0^{\infty}\frac{\sin( r k)}{ r^2}\prod\limits_{j=1}^3\sin( r k_j) d r =\\
&\frac{8\pi^3}{kk_1k_2k_3}(-|k+k_1-k_2-k_3|-|k-k_1+k_2-k_3|+\\
&~~~~|k+k_1+k_2-k_3|-|k-k_1-k_2+k_3|+\\
&~~~~|k+k_1-k_2+k_3|+ |k-k_1+k_2+k_3|+\\
&~~~|-k+k_1+k_2+k_3|-|k+k_1+k_2+k_3|).    
\end{aligned}
\end{equation*}
This expression can be considerably simplified after 
taking into account the four-wave frequency resonance condition
$$
k^2+k_1^2=k_2^2+k_3^2.
$$
This leads to
$$
\mathcal{P}=\frac{32\pi^3}{kk_1k_2k_3}\min(k,k_1,k_2,k_3).
$$
Substituting this into \eqref{WKEav}, we obtain the angle-averaged WKE:
\begin{widetext}
\begin{equation}
\label{WKEav2}
\frac{dn_k}{dt} = \frac{32\pi^3}{k}\int\min(k,k_1,k_2,k_3)\delta_{1\omega}^{23}n_kn_1n_2n_3 \bigg(\frac{1}{n_k}+\frac{1}{n_1}-\frac{1}{n_2}-\frac{1}{n_3}\bigg)k_1k_2k_3dk_1 dk_2 dk_3.
\end{equation}
\end{widetext}

The last point, which we shall address here, is passage in \eqref{WKEav2} to the frequency variables $\omega=k^2$ and $\omega_j=k_j^2$, $j=1,2,3$. Using the expressions $k=\sqrt{\omega}$, $dk=\frac{1}{2}\omega^{-1/2}$, $k_j=\sqrt{\omega_j}$, $dk_j=\frac{1}{2}\omega_j^{-1/2}$ and denoting $n_{\omega}=n_k$ and $$S(\omega,\omega_1,\omega_2,\omega_3) = \min(\sqrt{\omega},\sqrt{\omega_1},\sqrt{\omega_2},\sqrt{\omega_3}),$$
one obtains \eqref{E01} from \eqref{WKEav2}.

\section{Extra numerical details for WKE}
\label{Numerics}

\begin{figure}[hbt!]
\includegraphics[scale=0.35]{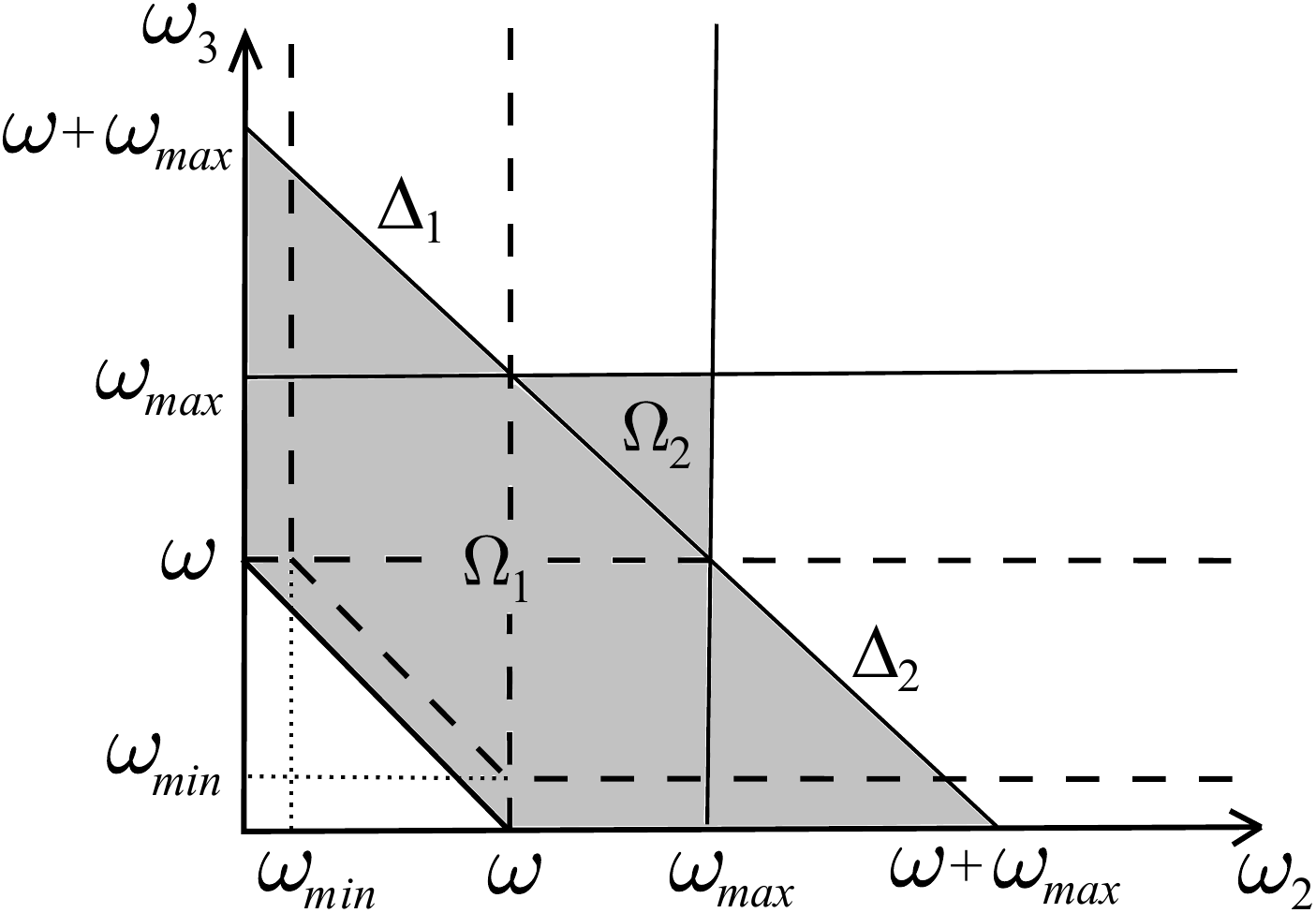}%
\caption{\label{domain}\textcolor{black}{Domain of integration $\Delta_{\omega}$ in the plain $(\omega_2,\omega_3)$ and its decomposition.}}
\end{figure}

An important block of the algorithm for solving WKE, which should be mentioned first, is an accurate and fast computation of the collision integral in the RHS of \eqref{E1}. To perform such a computation, the cubature formulae proposed in \cite{SemGreMedNaz} are applied, which requires \textcolor{black}{to decompose} the domain of integration $\Delta_{\omega}$ in \eqref{eta}, \eqref{gam}. In our study $\Delta_{\omega}$ is represented as $\Omega_1\cup\Omega_2$ \textcolor{black}{(gray area in Figure~\ref{domain})}, where \\

$\Omega_1=\{(\omega_2,\omega_3):\omega_2,\omega_3\geq 0,~0\leq\omega_2+\omega_3\textcolor{black}{-\omega}\leq \omega_{max}\},$\\

$\Omega_2 = \{(\omega_2,\omega_3):\omega_2,\omega_3\leq\omega_{max},~\omega_2+\omega_3\textcolor{black}{-\omega\geq} \omega_{max}\}.$

The integrands of RHSs of \eqref{eta}, \eqref{gam} have singularities along the lines $\omega_2=\omega$, $\omega_3=\omega$, $\omega_2=\omega_{min}$, $\omega_3=\omega_{min}$, $\omega_2+\omega_3\textcolor{black}{-\omega}=\omega_{min}$ \textcolor{black}{denoted by dashe lines in Figure~\ref{domain} },
where the first and the higher order derivatives of the integrands can have discontinuities. Therefore, to design cubature formulae with an exponential rate of convergence, $\Delta_{\omega}$ must be cut along these lines and decomposed into triangular, rectangular, and trapezoidal subdomain\textcolor{black}{, as shown in Figure~\ref{domain}}. Inside each subdomain the integrands are infinitely differentiable functions, see \cite{SemGreMedNaz} and \textcolor{black}{the reasoning presented} there. Each of the subdomains is mapped onto the reference square $[-1,1]^2$ (see \cite{Hossain} for the additional details), and tensor products of the modified Clenshaw--Curtis formulae are used to compute the integrals. 
It should be noted that, in contrast to the formulation considered in \cite{SemGreMedNaz}, here we have a bounded domain of integration \textcolor{black}{(gray area in Figure~\ref{domain})}, which somewhat simplifies the problem. \textcolor{black}{The contributions to the collision term from the triangles $\Delta_1$, $\Delta_2$ are the integrals of the terms $S_{\omega}^{23}n_c n_3$ and $S_{\omega}^{23}n_c n_2$  in \eqref{gam} respectively. The contribution from the domain $\Omega_2$ is the integral of $S_{\omega}^{23}n_2n_3$. Integration over other subdomains of $\Delta_{\omega}$ include all the terms  in \eqref{eta}, \eqref{gam}. } 

Now, let us address the problem of approximating the solution of WKE, $n_{\omega}(t)$. Here, first of all one should choose an appropriate interval of approximation with respect to the variable $\omega$. Since the spectrum vanishes at large $\omega$ \textcolor{black}{very rapidly}, it is enough to build approximation on the finite interval $[\omega_{min}, \omega_{max}]$ with large enough $\omega_{max}$ and set $n_{\omega}(t)=0$ for any $t<T$ and $\omega>\omega_{max}$. One should also take into account that the solution of WKE  blows up at the zero mode at some finite time $t^*$. Moreover, in the vicinity of $t=t^*$ the spectrum demonstrates a self-similar behavior with constant value at small  $\omega$, see \cite{semikoz1995kinetics, connaughton2004kinetic, lacaze2001dynamical, SemGreMedNaz} for details. To exclude the singularity at $\omega=0$, we choose a small positive $\omega_{min}$ and \textcolor{black}{fix the value $n_{\omega}(t) = n_{\omega_{min}}(t)$ for all $\omega<\omega_{min}$ at a given time moment $t<T$. This value is included in the integral over the thin layer between the dashed line and the outer boundary of $\Delta_{\omega}$ (see Figure~\ref{domain}).} Here, following the above results we set $T=100$.

For approximating WKE with respect to the time variable, we use the Runge--Kutta method of the order 2 (RK2), introduce the uniform grid with the step $dt$ and the nodes $t_j=jdt$, $j=0,1,...$ and implement an explicit time-marching scheme. Then, we denote $n^{[j]}=n^{[j]}(\omega)=n_{\omega}(t_j)$ and approximate the function $n^{[j]}(\omega)$ on the segment $[\omega_{min},\omega_{max}]$ by the interpolation polynomial with Chebyshev nodes. For this we use the barycentric interpolation formula from \cite{bar}:
  \begin{equation}
  \label{bar}
  p_M[n^{[j]}](\omega)=\sum\limits_{m=1}^M\frac{\xi_m n^{[j]}(\omega_m)}{\mathcal{L}(\omega)-y_m}\bigg/\sum\limits_{m=1}^M\frac{\xi_m}{\mathcal{L}(\omega)-y_m},
  \end{equation}
  where
  $\xi_m=\frac{1}{T_M'(y_m)}=(-1)^{m-1}\sin\bigl(\frac{2m-1}{2M}\pi\bigr)$
  are the weights of interpolation; $T_M(y)$ is the Chebyshev
  polynomial of the degree $M$; $y_m$ are zeroes of $T_M(y)$, $m=1,...,M$; $\mathcal{L}(\omega)$ is the linear mapping of the segment $[\omega_{min},\omega_{max}]$ to $[-1,1]$;
  $\omega_m=\mathcal{L}^{-1}(y_m)$ are the interpolation nodes.
  
  We also use the rational variant of \eqref{bar}, which turns out to be a very efficient way of adapting the approximation to singularities of the solution, see \cite{Bur,TeeTref}. In our case the singularities are related with a huge length of the considered $\omega$-interval (up to 7 decades) and with the blow-up of the solution at the zero mode. To obtain the rational barycentric interpolation, we fix the weights $\xi_m$ of \eqref{bar}, but instead of the Chebyshev nodes $y_m$ we use the transformed ones $\tilde{y}_m=g(y_m)$, $m=\overline{1,M}$. Here, the mapping $g:[-1,1]\to[-1,1]$ is bijection, which has an analytical continuation to the complex plain. This continuation must be reversible, and the inverse mapping $g^{-1}$ must take the singular point $\tilde{y}^*_1=\mathcal{L}(0)$ far enough from the segment $[-1,1]$. Several possibilities of defining $g(y)$ are discussed in \cite{Hoss, SemKuz, TeeTref}. However, the straightforward application of these formulae leads to very strong concentration of the nodes in the vicinity of $\omega=0$ and bad resolution of the initial Gaussian peak centered at $\omega=k_s^2$.

To take into account both singular points
$\tilde{y}^*_1=\mathcal{L}(0)$ and
$\tilde{y}^*_2=\mathcal{L}(k_s^2)$, we used the idea from \cite{Semen},
which allows one to write the mapping in the form
$$
g=\bigg(\frac{g_1^{-1}+g_2^{-1}}{2}\bigg)^{-1},
$$
where the superscript "$^{-1}$"\ means the inversion of a
function; $g_1^{-1}$, $g_2^{-1}$ serve for mapping the singular points
$\tilde{y}^*_1$ and $\tilde{y}^*_2$, respectively. In our case

$g_{i}^{-1}(\tilde{y}) = \dfrac{a^+-a^-+2\sinh^{-1}{\dfrac{\tilde{y}-\tilde{y}^*_i}{\epsilon_i}}}{a^++a^-},$\\

$a^{\pm} = \sinh^{-1}{\dfrac{1\pm\tilde{y}^*_i}{\epsilon_i}}$,

\noindent where $i=1,2$, $\epsilon_1$, $\epsilon_2$ are the small positive parameters
used for tuning the densities of concentration of nodes in the
vicinities of $\tilde{y}^*_1$ and $\tilde{y}^*_2$, respectively.
The closer $\epsilon_{1,2}$ are to zero, the higher is the density
of the concentration. Basically, we set $\epsilon_{1}=10^{-4}$, $\epsilon_2=10^{-1}$. 
It is worth noting that these parameters also have an important mathematical meaning. For the details about this issue we refer the reader to \cite{Hoss, TeeTref}. The described mapping $g$ in combination with the barycentric representation \eqref{bar} allows one to achieve the high accuracy even using coarse grids. However, it also requires for some modifications of the formulae for integrating the obtained spectra.

For computing the centroids by formulae \eqref{centroids} we first make the
change of the variable $k=\sqrt{\omega}$, then use the mapping
$\omega = \mathcal{L}^{-1}\circ g (y)$ with the Jacobian
$$
J_{\mathcal{L}g}(y) =
\frac{\omega_{max}-\omega_{min}}{1/(g_1'\circ g_1^{-1}\circ
g(y))+1/(g_2'\circ g_2^{-1}\circ
g(y))},
$$
and, finally, for integration with respect to the variable
$y=g^{-1}\circ\mathcal{L}(\omega)$ we used the Clenshaw--Curtis
formulae, see \cite{Clenshaw}.

In numerical tests we checked the convergence of the proposed method by observing the quantity
$$
R_{M}(n) = R_{M,t}(n) = \frac{\max\limits_{\omega}|n_M(\omega,t)-n_{2M}(\omega,t)|}{\max\limits_{\omega}|n_{2M}(\omega,t)|}
$$
obtained in computations with the fixed and small enough value of the time step $dt$, where $n_{M}(\omega,t)=p_M[n^{[j]}](\omega)$ is the rational interpolation of the spectrum with $M$ nodes, see \eqref{bar}. Here, it is assumed that for computing the collision integral we use the cubature formulae with $M^2$ nodes in each subdomain of $\Omega$. We also computed the quantities
$$
R_{M}(\ae) = R_{M,t}(\ae) = \frac{|\ae_M(t)-\ae_{2M}(t)|}{\ae_{2M}(t)},
$$
where $\ae $ denotes one of the functions $K_N(t)$, $K_E(t)$, $\Delta_{K_N}(t)$, $\Delta_{K_E}(t)$ determined in \eqref{centroids}, and the subscript $M$ stands for the number of nodes of spatial grid used for the approximation of the solution and for the number of nodes of all the applied formulae for numerical integration.

By analogy we computed the values
$$
R_{dt}(n) = R_{dt,t}(n) = \frac{\max\limits_{\omega}|n_{dt}(\omega,t)-n_{dt/2}(\omega,t)|}{\max\limits_{\omega}|n_{dt/2}(\omega,t)|}
$$
with the fixed and large enough number of space nodes, where $n_{dt}(\omega,t)$ is the solution obtained using RK2 with the time step $dt$. As well, we computed the values
$$
R_{dt}(\ae) = R_{dt,t}(\ae) = \frac{|\ae_{dt}(t)-\ae_{dt/2}(t)|}{|\ae_{dt/2}(t)|},
$$
where $\ae_{dt}(t)$ is determined similarly to $n_{dt}(\omega,t)$.

In Tables~\ref{conv1},~\ref{conv2} the values of $R_{M}$ and $R_{dt}$ are given. These values are obtained by running the test computations with the initial data \eqref{initialData} till $t=100$ and $t=50$, respectively.
\begin{table}[htb]
\caption{\label{conv1}%
Convergence of the algorithm for solving WKE  for the fixed $M=128$, $t=100$ and various values of the time step $dt$}
\begin{ruledtabular}
\begin{tabular}{cccccc}
$dt$ & $R_{dt}(n)$ & $R_{dt}(K_N)$ & $R_{dt}(\Delta_{K_N})$ & $R_{dt}(K_E)$ & $R_{dt}(\Delta_{K_E})$\\
\hline
2 & 6.75E-03 & 1.98E-03 & 1.03E-03 & 6.43E-05 &	9.51E-05\\
1 & 1.72E-03 & 4.99E-04 & 2.58E-04 & 1.58E-05 & 2.33E-05\\
1/2 & 4.32E-04 & 1.25E-04 & 6.45E-05 & 3.89E-06 & 5.75E-06\\
1/4 & 1.08E-04 & 3.12E-05 & 1.61E-05 & 9.65E-07 & 1.43E-06\\
1/8 & 2.71E-05 & 7.80E-06 & 4.02E-06 & 2.40E-07 & 3.55E-07\\
\end{tabular}
\end{ruledtabular}
\end{table}

\begin{table}[htb]
\caption{\label{conv2}%
Convergence of the algorithm for solving WKE for the fixed $dt=0.25$, $t=50$ and various values of the number $M$ of the nodes of spatial grids}
\begin{ruledtabular}
\begin{tabular}{cccccc}
$M$ & $R_{M}(n)$ & $R_{M}(K_N)$ & $R_{M}(\Delta_{K_N})$ & $R_{M}(K_E)$ & $R_{M}(\Delta_{K_E})$\\
\hline
32 & 3.61E-03 & 8.98E-04 & 8.37E-04 & 7.87E-04 & 3.73E-03\\
64 & 1.54E-05 & 4.37E-06 & 1.45E-06 & 7.98E-07 & 1.01E-05\\
128 & 1.85E-10 & 6.70E-09 & 1.50E-09 & 7.49E-11 & 4.94E-10\\
\end{tabular}
\end{ruledtabular}
\end{table}
We can state from these results that the rate of convergence with respect to time step is close to 2 that strictly corresponds to the theoretical estimates of the error of RK2. The convergence with respect to the number of nodes of the spatial grids is exponential, which also agrees with theoretical predictions and with the data from \cite{SemGreMedNaz}.

The described results are obtained for the times $t=100$ and $t=50$ that are rather far from the time $t^*\approx 126.7$, at which the solution of the WKE blows up. For the times close to $t^*$ most of approximations to the solution fail. Let us finalize our research by studding the convergence of the numerical solutions of the WKE and the conservation of the invariants, namely the density of particles and of energy, in the vicinity of $t^*$. 

 To perform this study, we expanded the segment $[\omega_{min},\omega_{max}]$ to $[0.001, 100^2]$, started with initial data \eqref{initialData} and ran the computations with $M=256$ and with manual adaptation of $dt$ till $t=t_0\approx 124.14$. The adaptation is done in order to decrease the relative deviations of invariants from their initial values as much as possible. At the point $t=t_0$ the deviations start to grow rapidly, and we use the solution at $t=t_0$ as an initial data for our analysis.
 
 First, we analyse the convergence with respect to the time step by observing the values of $R_{dt}(n)$ for the fixed $M=256$ at $t\approx125.7$. From the data in Table~\ref{conv3} one can see that the impact of the time-stepping error is rather small, and the order of convergence corresponds to theoretical value. 
\begin{table}[htb]
\caption{\label{conv3}%
Values of $R_{dt}(n)$ obtained with the fixed $M=256$ at $t\approx125.74$}
\begin{ruledtabular}
\begin{tabular}{ccccccc}
$dt$ & 0.16 & 0.08 & 0.04 & 0.02 & 0.01 & 0.005\\
\hline
$R_{dt}(n)$ & 0.0031 & 8.7E-04 & 1.9E-04 & 4.9E-05 & 1.2E-05 &	3.1E-06\\
\end{tabular}
\end{ruledtabular}
\end{table}

Second, we consider the convergence of approximations in $\omega$-variable. In Table~\ref{conv4} the values of $R_{M}(n)$ computed for $\omega\in[0.1^2,86^2]$ are given for different  $M$. The time step is fixed, $dt=0.08$. In this experiment we are focused on growth of the error in the vicinity of blow up time. Therefore, we observe several time moments close to $t^*$. 

\begin{table}[htb]
\caption{\label{conv4}%
Values of $R_M(n)$ obtained with the fixed step $dt=0.08$ at various times close to $t^*$}
\begin{ruledtabular}
\begin{tabular}{ccccc}
$M$ & $t\approx124.22$ & $t\approx124.7$ & $t\approx125.18$ & $t\approx 125.74$\\
\hline
64 & 0.0199 & 0.1365 & 0.2592 & 0.4343 \\
128 & 0.0199 & 0.1316 & 0.2036 & 0.2677 \\
256 & 0.0155 & 0.1007 & 0.1982 & 0.3543 \\
512 & 0.0046 & 0.0365 & 0.0483 & 0.1458 \\
\end{tabular}
\end{ruledtabular}
\end{table}

For the numerical values of spectra obtained in the case $M=1024$ we compute the relative deviations of the particle and energy densities from their initial values. They are denoted by $D_N(t)$ and $D_H(t)$, respectively, and presented in Table~\ref{conv5}. The formulae for them are
$$
D_N(t)=\frac{N(t)-N(0)}{N(0)},~~~D_H(t)=\frac{H(t)-H(0)}{H(0)},
$$
where $N(t)$ and $H(t)$ are computed by substituting the numerical solution of the WKE obtained for the time moment $t$ into \eqref{pa}, \eqref{en}.
\begin{table}[htb]
\caption{\label{conv5}%
Relative deviation of $D_N(t)$ and $D_H(t)$ at various times close to $t^*$ obtained in computations with $M=1024$, $dt=0.08$}
\begin{ruledtabular}
\begin{tabular}{ccccc}
 & $t\approx124.22$ & $t\approx124.7$ & $t\approx125.18$ & $125.74$\\
\hline
$D_N(t)$  & 8.69E-05 & 7.61E-05 & 3.88E-05 & -3.705E-04 \\
$D_H(t)$  & -1.86E-04 & -1.96E-04 & -2.12E-04 & -1.87E-04 \\
\end{tabular}
\end{ruledtabular}
\end{table}

From the presented results one can conclude that the error of the numerical solution grows rapidly in the vicinity of blow up time, and even the usage of rational approximations with huge number of nodes does not allow us to come very close to $t^*$. The last more or less adequate approximation of the solution can be obtained at $t\approx 125.7$. The possible reason is that we used rational approximations only for the spectrum, not for the integrands of the collision terms, which obviously also have strong singularities in the vicinity of $t^*$. Further developments will be directed to design of new cubature formulae aimed at the improvement of this drawback. 

An interesting fact is that rather large deviations of the numerical solutions reported in the last column of Table~\ref{conv4} does not affect so much the conservation of invariants. In simulation with $M=1024$ the value of $|D_N|$ jumps a bit to the order of $10^{-4}$, and $|D_H|$ remains practically unchanged. The reason is that the largest deviations of numerical solutions in tests with different $M$, and hence the errors, are localized in vicinity of the point $\omega_{min}$, whereas in all other points of the domain $[\omega_{min}, \omega_{max}]$ they remain relatively small. For example, comparison of $512$ and $1024$ results on the segment $[1, \omega_{max}]$ gives the relative deviation less than~$10^{-3}$.

While comparing the non-conservation of particle and energy densities, one should bring in mind the definitions \eqref{pa}, \eqref{en}. Due to the factor $\omega^{1/2}$ in \eqref{pa} the density of particles is more sensitive to the error of spectrum in the vicinity of $\omega_{min}$ than the density on energy. That is why for the times close to $t^*$ the deviation $D_N(t)$ starts to grow first.

\bibliography{apssamp}

\begin{thebibliography}{56}%
\makeatletter
\providecommand \@ifxundefined [1]{%
 \@ifx{#1\undefined}
}%
\providecommand \@ifnum [1]{%
 \ifnum #1\expandafter \@firstoftwo
 \else \expandafter \@secondoftwo
 \fi
}%
\providecommand \@ifx [1]{%
 \ifx #1\expandafter \@firstoftwo
 \else \expandafter \@secondoftwo
 \fi
}%
\providecommand \natexlab [1]{#1}%
\providecommand \enquote  [1]{``#1''}%
\providecommand \bibnamefont  [1]{#1}%
\providecommand \bibfnamefont [1]{#1}%
\providecommand \citenamefont [1]{#1}%
\providecommand \href@noop [0]{\@secondoftwo}%
\providecommand \href [0]{\begingroup \@sanitize@url \@href}%
\providecommand \@href[1]{\@@startlink{#1}\@@href}%
\providecommand \@@href[1]{\endgroup#1\@@endlink}%
\providecommand \@sanitize@url [0]{\catcode `\\12\catcode `\$12\catcode
  `\&12\catcode `\#12\catcode `\^12\catcode `\_12\catcode `\%12\relax}%
\providecommand \@@startlink[1]{}%
\providecommand \@@endlink[0]{}%
\providecommand \url  [0]{\begingroup\@sanitize@url \@url }%
\providecommand \@url [1]{\endgroup\@href {#1}{\urlprefix }}%
\providecommand \urlprefix  [0]{URL }%
\providecommand \Eprint [0]{\href }%
\providecommand \doibase [0]{https://doi.org/}%
\providecommand \selectlanguage [0]{\@gobble}%
\providecommand \bibinfo  [0]{\@secondoftwo}%
\providecommand \bibfield  [0]{\@secondoftwo}%
\providecommand \translation [1]{[#1]}%
\providecommand \BibitemOpen [0]{}%
\providecommand \bibitemStop [0]{}%
\providecommand \bibitemNoStop [0]{.\EOS\space}%
\providecommand \EOS [0]{\spacefactor3000\relax}%
\providecommand \BibitemShut  [1]{\csname bibitem#1\endcsname}%
\let\auto@bib@innerbib\@empty
\bibitem [{\citenamefont {Nazarenko}(2011)}]{nazarenko2011wave}%
  \BibitemOpen
  \bibfield  {author} {\bibinfo {author} {\bibfnamefont {S.}~\bibnamefont
  {Nazarenko}},\ }\href@noop {} {\emph {\bibinfo {title} {Wave turbulence}}},\
  Vol.\ \bibinfo {volume} {825}\ (\bibinfo  {publisher} {Springer Science \&
  Business Media},\ \bibinfo {year} {2011})\BibitemShut {NoStop}%
\bibitem [{\citenamefont {Zakharov}\ \emph {et~al.}(1992)\citenamefont
  {Zakharov}, \citenamefont {L'vov},\ and\ \citenamefont {Falkovich}}]{ZLF}%
  \BibitemOpen
  \bibfield  {author} {\bibinfo {author} {\bibfnamefont {V.~E.}\ \bibnamefont
  {Zakharov}}, \bibinfo {author} {\bibfnamefont {V.~S.}\ \bibnamefont
  {L'vov}},\ and\ \bibinfo {author} {\bibfnamefont {G.}~\bibnamefont
  {Falkovich}},\ }\bibfield  {title} {\bibinfo {title} {{Kolmogorov spectra of
  turbulence 1: Wave turbulence}},\ }\href@noop {} {\bibfield  {journal}
  {\bibinfo  {journal} {Springer Series in Nonlinear Dynamics}\ } (\bibinfo
  {year} {1992})}\BibitemShut {NoStop}%
\bibitem [{\citenamefont {Dyachenko}\ \emph {et~al.}(1992)\citenamefont
  {Dyachenko}, \citenamefont {Newell}, \citenamefont {Pushkarev},\ and\
  \citenamefont {Zakharov}}]{dyachenko1992optical}%
  \BibitemOpen
  \bibfield  {author} {\bibinfo {author} {\bibfnamefont {S.}~\bibnamefont
  {Dyachenko}}, \bibinfo {author} {\bibfnamefont {A.}~\bibnamefont {Newell}},
  \bibinfo {author} {\bibfnamefont {A.}~\bibnamefont {Pushkarev}},\ and\
  \bibinfo {author} {\bibfnamefont {V.}~\bibnamefont {Zakharov}},\ }\bibfield
  {title} {\bibinfo {title} {{Optical turbulence: weak turbulence, condensates
  and collapsing filaments in the nonlinear Schr{\"o}dinger equation}},\
  }\href@noop {} {\bibfield  {journal} {\bibinfo  {journal} {Physica D:
  Nonlinear Phenomena}\ }\textbf {\bibinfo {volume} {57}},\ \bibinfo {pages}
  {96} (\bibinfo {year} {1992})}\BibitemShut {NoStop}%
\bibitem [{\citenamefont {Proment}\ \emph {et~al.}(2009)\citenamefont
  {Proment}, \citenamefont {Nazarenko},\ and\ \citenamefont
  {Onorato}}]{proment2009energy}%
  \BibitemOpen
  \bibfield  {author} {\bibinfo {author} {\bibfnamefont {D.}~\bibnamefont
  {Proment}}, \bibinfo {author} {\bibfnamefont {S.}~\bibnamefont {Nazarenko}},\
  and\ \bibinfo {author} {\bibfnamefont {M.}~\bibnamefont {Onorato}},\
  }\bibfield  {title} {\bibinfo {title} {Quantum turbulence cascades in the
  gross-pitaevskii model},\ }\href {https://doi.org/10.1103/PhysRevA.80.051603}
  {\bibfield  {journal} {\bibinfo  {journal} {Phys. Rev. A}\ }\textbf {\bibinfo
  {volume} {80}},\ \bibinfo {pages} {051603} (\bibinfo {year}
  {2009})}\BibitemShut {NoStop}%
\bibitem [{\citenamefont {Proment}\ \emph
  {et~al.}(2012{\natexlab{a}})\citenamefont {Proment}, \citenamefont
  {Nazarenko},\ and\ \citenamefont {Onorato}}]{proment2012sustained}%
  \BibitemOpen
  \bibfield  {author} {\bibinfo {author} {\bibfnamefont {D.}~\bibnamefont
  {Proment}}, \bibinfo {author} {\bibfnamefont {S.}~\bibnamefont {Nazarenko}},\
  and\ \bibinfo {author} {\bibfnamefont {M.}~\bibnamefont {Onorato}},\
  }\bibfield  {title} {\bibinfo {title} {Sustained turbulence in the
  three-dimensional gross--pitaevskii model},\ }\href@noop {} {\bibfield
  {journal} {\bibinfo  {journal} {Physica D: nonlinear phenomena}\ }\textbf
  {\bibinfo {volume} {241}},\ \bibinfo {pages} {304} (\bibinfo {year}
  {2012}{\natexlab{a}})}\BibitemShut {NoStop}%
\bibitem [{\citenamefont {Navon}\ \emph {et~al.}(2016)\citenamefont {Navon},
  \citenamefont {Gaunt}, \citenamefont {Smith},\ and\ \citenamefont
  {Hadzibabic}}]{navon2016emergence}%
  \BibitemOpen
  \bibfield  {author} {\bibinfo {author} {\bibfnamefont {N.}~\bibnamefont
  {Navon}}, \bibinfo {author} {\bibfnamefont {A.~L.}\ \bibnamefont {Gaunt}},
  \bibinfo {author} {\bibfnamefont {R.~P.}\ \bibnamefont {Smith}},\ and\
  \bibinfo {author} {\bibfnamefont {Z.}~\bibnamefont {Hadzibabic}},\ }\bibfield
   {title} {\bibinfo {title} {Emergence of a turbulent cascade in a quantum
  gas},\ }\href@noop {} {\bibfield  {journal} {\bibinfo  {journal} {Nature}\
  }\textbf {\bibinfo {volume} {539}},\ \bibinfo {pages} {72} (\bibinfo {year}
  {2016})}\BibitemShut {NoStop}%
\bibitem [{\citenamefont {Navon}\ \emph {et~al.}(2019)\citenamefont {Navon},
  \citenamefont {Eigen}, \citenamefont {Zhang}, \citenamefont {Lopes},
  \citenamefont {Gaunt}, \citenamefont {Fujimoto}, \citenamefont {Tsubota},
  \citenamefont {Smith},\ and\ \citenamefont
  {Hadzibabic}}]{navon2019synthetic}%
  \BibitemOpen
  \bibfield  {author} {\bibinfo {author} {\bibfnamefont {N.}~\bibnamefont
  {Navon}}, \bibinfo {author} {\bibfnamefont {C.}~\bibnamefont {Eigen}},
  \bibinfo {author} {\bibfnamefont {J.}~\bibnamefont {Zhang}}, \bibinfo
  {author} {\bibfnamefont {R.}~\bibnamefont {Lopes}}, \bibinfo {author}
  {\bibfnamefont {A.~L.}\ \bibnamefont {Gaunt}}, \bibinfo {author}
  {\bibfnamefont {K.}~\bibnamefont {Fujimoto}}, \bibinfo {author}
  {\bibfnamefont {M.}~\bibnamefont {Tsubota}}, \bibinfo {author} {\bibfnamefont
  {R.~P.}\ \bibnamefont {Smith}},\ and\ \bibinfo {author} {\bibfnamefont
  {Z.}~\bibnamefont {Hadzibabic}},\ }\bibfield  {title} {\bibinfo {title}
  {Synthetic dissipation and cascade fluxes in a turbulent quantum gas},\
  }\href@noop {} {\bibfield  {journal} {\bibinfo  {journal} {Science}\ }\textbf
  {\bibinfo {volume} {366}},\ \bibinfo {pages} {382} (\bibinfo {year}
  {2019})}\BibitemShut {NoStop}%
\bibitem [{\citenamefont {Semikoz}\ and\ \citenamefont
  {Tkachev}(1995)}]{semikoz1995kinetics}%
  \BibitemOpen
  \bibfield  {author} {\bibinfo {author} {\bibfnamefont {D.~V.}\ \bibnamefont
  {Semikoz}}\ and\ \bibinfo {author} {\bibfnamefont {I.~I.}\ \bibnamefont
  {Tkachev}},\ }\bibfield  {title} {\bibinfo {title} {{Kinetics of Bose
  condensation}},\ }\href@noop {} {\bibfield  {journal} {\bibinfo  {journal}
  {Physical review letters}\ }\textbf {\bibinfo {volume} {74}},\ \bibinfo
  {pages} {3093} (\bibinfo {year} {1995})}\BibitemShut {NoStop}%
\bibitem [{\citenamefont {Lacaze}\ \emph {et~al.}(2001)\citenamefont {Lacaze},
  \citenamefont {Lallemand}, \citenamefont {Pomeau},\ and\ \citenamefont
  {Rica}}]{lacaze2001dynamical}%
  \BibitemOpen
  \bibfield  {author} {\bibinfo {author} {\bibfnamefont {R.}~\bibnamefont
  {Lacaze}}, \bibinfo {author} {\bibfnamefont {P.}~\bibnamefont {Lallemand}},
  \bibinfo {author} {\bibfnamefont {Y.}~\bibnamefont {Pomeau}},\ and\ \bibinfo
  {author} {\bibfnamefont {S.}~\bibnamefont {Rica}},\ }\bibfield  {title}
  {\bibinfo {title} {{Dynamical formation of a Bose--Einstein condensate}},\
  }\href@noop {} {\bibfield  {journal} {\bibinfo  {journal} {Physica D:
  Nonlinear Phenomena}\ }\textbf {\bibinfo {volume} {152}},\ \bibinfo {pages}
  {779} (\bibinfo {year} {2001})}\BibitemShut {NoStop}%
\bibitem [{\citenamefont {Connaughton}\ and\ \citenamefont
  {Pomeau}(2004)}]{connaughton2004kinetic}%
  \BibitemOpen
  \bibfield  {author} {\bibinfo {author} {\bibfnamefont {C.}~\bibnamefont
  {Connaughton}}\ and\ \bibinfo {author} {\bibfnamefont {Y.}~\bibnamefont
  {Pomeau}},\ }\bibfield  {title} {\bibinfo {title} {{Kinetic theory and
  Bose--Einstein condensation}},\ }\href@noop {} {\bibfield  {journal}
  {\bibinfo  {journal} {Comptes Rendus Physique}\ }\textbf {\bibinfo {volume}
  {5}},\ \bibinfo {pages} {91} (\bibinfo {year} {2004})}\BibitemShut {NoStop}%
\bibitem [{\citenamefont {Bell}\ \emph {et~al.}(2017)\citenamefont {Bell},
  \citenamefont {Grebenev}, \citenamefont {Medvedev},\ and\ \citenamefont
  {Nazarenko}}]{BellNaz}%
  \BibitemOpen
  \bibfield  {author} {\bibinfo {author} {\bibfnamefont {N.}~\bibnamefont
  {Bell}}, \bibinfo {author} {\bibfnamefont {V.}~\bibnamefont {Grebenev}},
  \bibinfo {author} {\bibfnamefont {S.}~\bibnamefont {Medvedev}},\ and\
  \bibinfo {author} {\bibfnamefont {S.}~\bibnamefont {Nazarenko}},\ }\bibfield
  {title} {\bibinfo {title} {{Self-similar evolution of Alfven wave
  turbulence}},\ }\href@noop {} {\bibfield  {journal} {\bibinfo  {journal}
  {Journal of Physics A: Mathematical and Theoretical}\ }\textbf {\bibinfo
  {volume} {50}},\ \bibinfo {pages} {435501} (\bibinfo {year}
  {2017})}\BibitemShut {NoStop}%
\bibitem [{\citenamefont {Bell}\ and\ \citenamefont
  {Nazarenko}(2018)}]{Bell_2018}%
  \BibitemOpen
  \bibfield  {author} {\bibinfo {author} {\bibfnamefont {N.~K.}\ \bibnamefont
  {Bell}}\ and\ \bibinfo {author} {\bibfnamefont {S.~V.}\ \bibnamefont
  {Nazarenko}},\ }\bibfield  {title} {\bibinfo {title} {{Reflected wave
  solution of Alfv{\'{e}}n wave turbulence}},\ }\href
  {https://doi.org/10.1088/1751-8121/aad833} {\bibfield  {journal} {\bibinfo
  {journal} {Journal of Physics A: Mathematical and Theoretical}\ }\textbf
  {\bibinfo {volume} {51}},\ \bibinfo {pages} {405501} (\bibinfo {year}
  {2018})}\BibitemShut {NoStop}%
\bibitem [{\citenamefont {Semisalov}\ \emph {et~al.}(2021)\citenamefont
  {Semisalov}, \citenamefont {Grebenev}, \citenamefont {Medvedev},\ and\
  \citenamefont {Nazarenko}}]{SemGreMedNaz}%
  \BibitemOpen
  \bibfield  {author} {\bibinfo {author} {\bibfnamefont {B.}~\bibnamefont
  {Semisalov}}, \bibinfo {author} {\bibfnamefont {V.}~\bibnamefont {Grebenev}},
  \bibinfo {author} {\bibfnamefont {S.}~\bibnamefont {Medvedev}},\ and\
  \bibinfo {author} {\bibfnamefont {S.}~\bibnamefont {Nazarenko}},\ }\bibfield
  {title} {\bibinfo {title} {{Numerical analysis of a self-similar turbulent
  flow in Bose–-Einstein condensates}},\ }\href
  {https://doi.org/https://doi.org/10.1016/j.cnsns.2021.105903} {\bibfield
  {journal} {\bibinfo  {journal} {Communications in Nonlinear Science and
  Numerical Simulation}\ ,\ \bibinfo {pages} {105903}} (\bibinfo {year}
  {2021})}\BibitemShut {NoStop}%
\bibitem [{\citenamefont {Lvov}\ and\ \citenamefont
  {Nazarenko}(2004)}]{lvov_2004}%
  \BibitemOpen
  \bibfield  {author} {\bibinfo {author} {\bibfnamefont {Y.~V.}\ \bibnamefont
  {Lvov}}\ and\ \bibinfo {author} {\bibfnamefont {S.}~\bibnamefont
  {Nazarenko}},\ }\bibfield  {title} {\bibinfo {title} {Noisy spectra, long
  correlations, and intermittency in wave turbulence},\ }\href
  {https://doi.org/10.1103/PhysRevE.69.066608} {\bibfield  {journal} {\bibinfo
  {journal} {Phys. Rev. E}\ }\textbf {\bibinfo {volume} {69}},\ \bibinfo
  {pages} {066608} (\bibinfo {year} {2004})}\BibitemShut {NoStop}%
\bibitem [{\citenamefont {Choi}\ \emph {et~al.}(2004)\citenamefont {Choi},
  \citenamefont {Lvov},\ and\ \citenamefont {Nazarenko}}]{choi_2004}%
  \BibitemOpen
  \bibfield  {author} {\bibinfo {author} {\bibfnamefont {Y.}~\bibnamefont
  {Choi}}, \bibinfo {author} {\bibfnamefont {Y.~V.}\ \bibnamefont {Lvov}},\
  and\ \bibinfo {author} {\bibfnamefont {S.}~\bibnamefont {Nazarenko}},\
  }\bibfield  {title} {\bibinfo {title} {Probability densities and preservation
  of randomness in wave turbulence},\ }\href
  {https://doi.org/https://doi.org/10.1016/j.physleta.2004.09.062} {\bibfield
  {journal} {\bibinfo  {journal} {Physics Letters A}\ }\textbf {\bibinfo
  {volume} {332}},\ \bibinfo {pages} {230} (\bibinfo {year}
  {2004})}\BibitemShut {NoStop}%
\bibitem [{\citenamefont {Choi}\ \emph
  {et~al.}(2005{\natexlab{a}})\citenamefont {Choi}, \citenamefont {Lvov},
  \citenamefont {Nazarenko},\ and\ \citenamefont {Pokorni}}]{choi_2005}%
  \BibitemOpen
  \bibfield  {author} {\bibinfo {author} {\bibfnamefont {Y.}~\bibnamefont
  {Choi}}, \bibinfo {author} {\bibfnamefont {Y.~V.}\ \bibnamefont {Lvov}},
  \bibinfo {author} {\bibfnamefont {S.}~\bibnamefont {Nazarenko}},\ and\
  \bibinfo {author} {\bibfnamefont {B.}~\bibnamefont {Pokorni}},\ }\bibfield
  {title} {\bibinfo {title} {Anomalous probability of large amplitudes in wave
  turbulence},\ }\href
  {https://doi.org/https://doi.org/10.1016/j.physleta.2005.02.072} {\bibfield
  {journal} {\bibinfo  {journal} {Physics Letters A}\ }\textbf {\bibinfo
  {volume} {339}},\ \bibinfo {pages} {361} (\bibinfo {year}
  {2005}{\natexlab{a}})}\BibitemShut {NoStop}%
\bibitem [{\citenamefont {Choi}\ \emph
  {et~al.}(2005{\natexlab{b}})\citenamefont {Choi}, \citenamefont {Lvov},\ and\
  \citenamefont {Nazarenko}}]{choi_2005b}%
  \BibitemOpen
  \bibfield  {author} {\bibinfo {author} {\bibfnamefont {Y.}~\bibnamefont
  {Choi}}, \bibinfo {author} {\bibfnamefont {Y.~V.}\ \bibnamefont {Lvov}},\
  and\ \bibinfo {author} {\bibfnamefont {S.}~\bibnamefont {Nazarenko}},\
  }\bibfield  {title} {\bibinfo {title} {Joint statistics of amplitudes and
  phases in wave turbulence},\ }\href
  {https://doi.org/https://doi.org/10.1016/j.physd.2004.11.016} {\bibfield
  {journal} {\bibinfo  {journal} {Physica D: Nonlinear Phenomena}\ }\textbf
  {\bibinfo {volume} {201}},\ \bibinfo {pages} {121} (\bibinfo {year}
  {2005}{\natexlab{b}})}\BibitemShut {NoStop}%
\bibitem [{\citenamefont {Shrira}\ and\ \citenamefont
  {Nazarenko}(2013)}]{shrira2013advances}%
  \BibitemOpen
  \bibfield  {author} {\bibinfo {author} {\bibfnamefont {V.}~\bibnamefont
  {Shrira}}\ and\ \bibinfo {author} {\bibfnamefont {S.}~\bibnamefont
  {Nazarenko}},\ }\href {https://books.google.es/books?id=pTm7CgAAQBAJ} {\emph
  {\bibinfo {title} {{Advances In Wave Turbulence}}}},\ World Scientific Series
  On Nonlinear Science Series A\ (\bibinfo  {publisher} {World Scientific
  Publishing Company},\ \bibinfo {year} {2013})\BibitemShut {NoStop}%
\bibitem [{\citenamefont {Deng}\ and\ \citenamefont {Hani}(2021)}]{hani}%
  \BibitemOpen
  \bibfield  {author} {\bibinfo {author} {\bibfnamefont {Y.}~\bibnamefont
  {Deng}}\ and\ \bibinfo {author} {\bibfnamefont {Z.}~\bibnamefont {Hani}},\
  }\bibfield  {title} {\bibinfo {title} {Full derivation of the wave kinetic
  equation},\ }\href@noop {} {\bibfield  {journal} {\bibinfo  {journal} {arXiv
  preprint arXiv:2106.0981}\ } (\bibinfo {year} {2021})}\BibitemShut {NoStop}%
\bibitem [{\citenamefont {Tanaka}\ and\ \citenamefont
  {Yokoyama}(2013)}]{Tanaka}%
  \BibitemOpen
  \bibfield  {author} {\bibinfo {author} {\bibfnamefont {M.}~\bibnamefont
  {Tanaka}}\ and\ \bibinfo {author} {\bibfnamefont {N.}~\bibnamefont
  {Yokoyama}},\ }\bibfield  {title} {\bibinfo {title} {Numerical verification
  of the random-phase-and-amplitude formalism of weak turbulence},\ }\href
  {https://doi.org/10.1103/PhysRevE.87.062922} {\bibfield  {journal} {\bibinfo
  {journal} {Phys. Rev. E}\ }\textbf {\bibinfo {volume} {87}},\ \bibinfo
  {pages} {062922} (\bibinfo {year} {2013})}\BibitemShut {NoStop}%
\bibitem [{\citenamefont {Choi}\ \emph {et~al.}(2017)\citenamefont {Choi},
  \citenamefont {Jo}, \citenamefont {Kwon},\ and\ \citenamefont
  {Nazarenko}}]{Choi_2017}%
  \BibitemOpen
  \bibfield  {author} {\bibinfo {author} {\bibfnamefont {Y.}~\bibnamefont
  {Choi}}, \bibinfo {author} {\bibfnamefont {S.}~\bibnamefont {Jo}}, \bibinfo
  {author} {\bibfnamefont {Y.-S.}\ \bibnamefont {Kwon}},\ and\ \bibinfo
  {author} {\bibfnamefont {S.}~\bibnamefont {Nazarenko}},\ }\bibfield  {title}
  {\bibinfo {title} {Nonstationary distributions of wave intensities in wave
  turbulence},\ }\href {https://doi.org/10.1088/1751-8121/aa7dba} {\bibfield
  {journal} {\bibinfo  {journal} {Journal of Physics A: Mathematical and
  Theoretical}\ }\textbf {\bibinfo {volume} {50}},\ \bibinfo {pages} {355502}
  (\bibinfo {year} {2017})}\BibitemShut {NoStop}%
\bibitem [{\citenamefont {Zakharov}\ \emph {et~al.}(2007)\citenamefont
  {Zakharov}, \citenamefont {Korotkevich}, \citenamefont {Pushkarev},\ and\
  \citenamefont {Resio}}]{PhysRevLett.99.164501}%
  \BibitemOpen
  \bibfield  {author} {\bibinfo {author} {\bibfnamefont {V.~E.}\ \bibnamefont
  {Zakharov}}, \bibinfo {author} {\bibfnamefont {A.~O.}\ \bibnamefont
  {Korotkevich}}, \bibinfo {author} {\bibfnamefont {A.}~\bibnamefont
  {Pushkarev}},\ and\ \bibinfo {author} {\bibfnamefont {D.}~\bibnamefont
  {Resio}},\ }\bibfield  {title} {\bibinfo {title} {Coexistence of weak and
  strong wave turbulence in a swell propagation},\ }\href
  {https://doi.org/10.1103/PhysRevLett.99.164501} {\bibfield  {journal}
  {\bibinfo  {journal} {Phys. Rev. Lett.}\ }\textbf {\bibinfo {volume} {99}},\
  \bibinfo {pages} {164501} (\bibinfo {year} {2007})}\BibitemShut {NoStop}%
\bibitem [{\citenamefont {Korotkevich}\ \emph {et~al.}(2008)\citenamefont
  {Korotkevich}, \citenamefont {Pushkarev}, \citenamefont {Resio},\ and\
  \citenamefont {Zakharov}}]{KOROTKEVICH}%
  \BibitemOpen
  \bibfield  {author} {\bibinfo {author} {\bibfnamefont {A.}~\bibnamefont
  {Korotkevich}}, \bibinfo {author} {\bibfnamefont {A.}~\bibnamefont
  {Pushkarev}}, \bibinfo {author} {\bibfnamefont {D.}~\bibnamefont {Resio}},\
  and\ \bibinfo {author} {\bibfnamefont {V.}~\bibnamefont {Zakharov}},\
  }\bibfield  {title} {\bibinfo {title} {Numerical verification of the weak
  turbulent model for swell evolution},\ }\href
  {https://doi.org/https://doi.org/10.1016/j.euromechflu.2007.08.004}
  {\bibfield  {journal} {\bibinfo  {journal} {European Journal of Mechanics -
  B/Fluids}\ }\textbf {\bibinfo {volume} {27}},\ \bibinfo {pages} {361}
  (\bibinfo {year} {2008})}\BibitemShut {NoStop}%
\bibitem [{\citenamefont {Korotkevich}\ \emph {et~al.}(2019)\citenamefont
  {Korotkevich}, \citenamefont {Prokofiev},\ and\ \citenamefont
  {Zakharov}}]{korotkevich2019dissipation}%
  \BibitemOpen
  \bibfield  {author} {\bibinfo {author} {\bibfnamefont {A.~O.}\ \bibnamefont
  {Korotkevich}}, \bibinfo {author} {\bibfnamefont {A.}~\bibnamefont
  {Prokofiev}},\ and\ \bibinfo {author} {\bibfnamefont {V.~E.}\ \bibnamefont
  {Zakharov}},\ }\bibfield  {title} {\bibinfo {title} {On the dissipation rate
  of ocean waves due to white capping},\ }\href@noop {} {\bibfield  {journal}
  {\bibinfo  {journal} {JETP Letters}\ }\textbf {\bibinfo {volume} {109}},\
  \bibinfo {pages} {309} (\bibinfo {year} {2019})}\BibitemShut {NoStop}%
\bibitem [{\citenamefont {Banks}\ \emph {et~al.}(2021)\citenamefont {Banks},
  \citenamefont {Buckmaster}, \citenamefont {Korotkevich}, \citenamefont
  {Kova{\v{c}}i{\v{c}}},\ and\ \citenamefont {Shatah}}]{banks2021direct}%
  \BibitemOpen
  \bibfield  {author} {\bibinfo {author} {\bibfnamefont {J.}~\bibnamefont
  {Banks}}, \bibinfo {author} {\bibfnamefont {T.}~\bibnamefont {Buckmaster}},
  \bibinfo {author} {\bibfnamefont {A.}~\bibnamefont {Korotkevich}}, \bibinfo
  {author} {\bibfnamefont {G.}~\bibnamefont {Kova{\v{c}}i{\v{c}}}},\ and\
  \bibinfo {author} {\bibfnamefont {J.}~\bibnamefont {Shatah}},\ }\bibfield
  {title} {\bibinfo {title} {Direct verification of the kinetic description of
  wave turbulence},\ }\href@noop {} {\bibfield  {journal} {\bibinfo  {journal}
  {arXiv preprint arXiv:2109.02477}\ } (\bibinfo {year} {2021})}\BibitemShut
  {NoStop}%
\bibitem [{\citenamefont {P{\'\i}tajevsk{\'\i}j}\ and\ \citenamefont
  {Stringari}(2003)}]{pitaevskii2003bose}%
  \BibitemOpen
  \bibfield  {author} {\bibinfo {author} {\bibfnamefont {L.}~\bibnamefont
  {P{\'\i}tajevsk{\'\i}j}}\ and\ \bibinfo {author} {\bibfnamefont
  {S.}~\bibnamefont {Stringari}},\ }\href
  {https://books.google.es/books?id=rIobbOxC4j4C} {\emph {\bibinfo {title}
  {{Bose--Einstein Condensation}}}},\ International Series of Monographs on
  Physics\ (\bibinfo  {publisher} {Clarendon Press},\ \bibinfo {year}
  {2003})\BibitemShut {NoStop}%
\bibitem [{\citenamefont {Fj{\o}rtoft}(1953)}]{fjortoft1953changes}%
  \BibitemOpen
  \bibfield  {author} {\bibinfo {author} {\bibfnamefont {R.}~\bibnamefont
  {Fj{\o}rtoft}},\ }\bibfield  {title} {\bibinfo {title} {On the changes in the
  spectral distribution of kinetic energy for twodimensional, nondivergent
  flow},\ }\href@noop {} {\bibfield  {journal} {\bibinfo  {journal} {Tellus}\
  }\textbf {\bibinfo {volume} {5}},\ \bibinfo {pages} {225} (\bibinfo {year}
  {1953})}\BibitemShut {NoStop}%
\bibitem [{\citenamefont {Zakharov}\ \emph {et~al.}(1985)\citenamefont
  {Zakharov}, \citenamefont {Musher},\ and\ \citenamefont
  {Rubenchik}}]{ZAKHAROV1985285}%
  \BibitemOpen
  \bibfield  {author} {\bibinfo {author} {\bibfnamefont {V.}~\bibnamefont
  {Zakharov}}, \bibinfo {author} {\bibfnamefont {S.}~\bibnamefont {Musher}},\
  and\ \bibinfo {author} {\bibfnamefont {A.}~\bibnamefont {Rubenchik}},\
  }\bibfield  {title} {\bibinfo {title} {Hamiltonian approach to the
  description of non-linear plasma phenomena},\ }\href
  {https://doi.org/https://doi.org/10.1016/0370-1573(85)90040-7} {\bibfield
  {journal} {\bibinfo  {journal} {Physics Reports}\ }\textbf {\bibinfo {volume}
  {129}},\ \bibinfo {pages} {285} (\bibinfo {year} {1985})}\BibitemShut
  {NoStop}%
\bibitem [{\citenamefont {Semikoz}\ and\ \citenamefont
  {Tkachev}(1997)}]{semikoz1997condensation}%
  \BibitemOpen
  \bibfield  {author} {\bibinfo {author} {\bibfnamefont {D.~V.}\ \bibnamefont
  {Semikoz}}\ and\ \bibinfo {author} {\bibfnamefont {I.~I.}\ \bibnamefont
  {Tkachev}},\ }\bibfield  {title} {\bibinfo {title} {Condensation of bosons in
  the kinetic regime},\ }\href@noop {} {\bibfield  {journal} {\bibinfo
  {journal} {Physical review D}\ }\textbf {\bibinfo {volume} {55}},\ \bibinfo
  {pages} {489} (\bibinfo {year} {1997})}\BibitemShut {NoStop}%
\bibitem [{\citenamefont {Proment}\ \emph
  {et~al.}(2012{\natexlab{b}})\citenamefont {Proment}, \citenamefont {Onorato},
  \citenamefont {Asinari},\ and\ \citenamefont {Nazarenko}}]{DMPS}%
  \BibitemOpen
  \bibfield  {author} {\bibinfo {author} {\bibfnamefont {D.}~\bibnamefont
  {Proment}}, \bibinfo {author} {\bibfnamefont {M.}~\bibnamefont {Onorato}},
  \bibinfo {author} {\bibfnamefont {P.}~\bibnamefont {Asinari}},\ and\ \bibinfo
  {author} {\bibfnamefont {S.}~\bibnamefont {Nazarenko}},\ }\bibfield  {title}
  {\bibinfo {title} {{Warm cascade states in a forced-dissipated Boltzmann gas
  of hard spheres}},\ }\href
  {https://doi.org/https://doi.org/10.1016/j.physd.2011.11.019} {\bibfield
  {journal} {\bibinfo  {journal} {Physica D: Nonlinear Phenomena}\ }\textbf
  {\bibinfo {volume} {241}},\ \bibinfo {pages} {600} (\bibinfo {year}
  {2012}{\natexlab{b}})}\BibitemShut {NoStop}%
\bibitem [{\citenamefont {Gottlieb}\ and\ \citenamefont
  {Orszag}(1977)}]{gottlieb1977numerical}%
  \BibitemOpen
  \bibfield  {author} {\bibinfo {author} {\bibfnamefont {D.}~\bibnamefont
  {Gottlieb}}\ and\ \bibinfo {author} {\bibfnamefont {S.~A.}\ \bibnamefont
  {Orszag}},\ }\href@noop {} {\emph {\bibinfo {title} {Numerical analysis of
  spectral methods: theory and applications}}}\ (\bibinfo  {publisher} {SIAM},\
  \bibinfo {year} {1977})\BibitemShut {NoStop}%
\bibitem [{\citenamefont {Frigo}\ and\ \citenamefont
  {Johnson}(2005)}]{frigo2005design}%
  \BibitemOpen
  \bibfield  {author} {\bibinfo {author} {\bibfnamefont {M.}~\bibnamefont
  {Frigo}}\ and\ \bibinfo {author} {\bibfnamefont {S.~G.}\ \bibnamefont
  {Johnson}},\ }\bibfield  {title} {\bibinfo {title} {The design and
  implementation of fftw3},\ }\href@noop {} {\bibfield  {journal} {\bibinfo
  {journal} {Proceedings of the IEEE}\ }\textbf {\bibinfo {volume} {93}},\
  \bibinfo {pages} {216} (\bibinfo {year} {2005})}\BibitemShut {NoStop}%
\bibitem [{\citenamefont {Krstulovic}(2020)}]{KrstulovicHDR}%
  \BibitemOpen
  \bibfield  {author} {\bibinfo {author} {\bibfnamefont {G.}~\bibnamefont
  {Krstulovic}},\ }\emph {\bibinfo {title} {{A theoretical description of
  vortex dynamics in superfluids. Kelvin waves, reconnections and
  particle-vortex interaction}}},\ \href
  {https://gkrstulovic.gitlab.io/thesisms/hdr-krstulovic/} {\bibinfo {type}
  {Habilitation \`a diriger des recherches}},\ \bibinfo  {school} {Universite
  C\^ote d’Azur} (\bibinfo {year} {2020})\BibitemShut {NoStop}%
\bibitem [{\citenamefont {Cox}\ and\ \citenamefont
  {Matthews}(2002)}]{cox2002exponential}%
  \BibitemOpen
  \bibfield  {author} {\bibinfo {author} {\bibfnamefont {S.~M.}\ \bibnamefont
  {Cox}}\ and\ \bibinfo {author} {\bibfnamefont {P.~C.}\ \bibnamefont
  {Matthews}},\ }\bibfield  {title} {\bibinfo {title} {Exponential time
  differencing for stiff systems},\ }\href@noop {} {\bibfield  {journal}
  {\bibinfo  {journal} {Journal of Computational Physics}\ }\textbf {\bibinfo
  {volume} {176}},\ \bibinfo {pages} {430} (\bibinfo {year}
  {2002})}\BibitemShut {NoStop}%
\bibitem [{\citenamefont {Connaughton}\ \emph {et~al.}(2001)\citenamefont
  {Connaughton}, \citenamefont {Nazarenko},\ and\ \citenamefont
  {Pushkarev}}]{PhysRevE.63.046306}%
  \BibitemOpen
  \bibfield  {author} {\bibinfo {author} {\bibfnamefont {C.}~\bibnamefont
  {Connaughton}}, \bibinfo {author} {\bibfnamefont {S.}~\bibnamefont
  {Nazarenko}},\ and\ \bibinfo {author} {\bibfnamefont {A.}~\bibnamefont
  {Pushkarev}},\ }\bibfield  {title} {\bibinfo {title} {Discreteness and
  quasiresonances in weak turbulence of capillary waves},\ }\href
  {https://doi.org/10.1103/PhysRevE.63.046306} {\bibfield  {journal} {\bibinfo
  {journal} {Phys. Rev. E}\ }\textbf {\bibinfo {volume} {63}},\ \bibinfo
  {pages} {046306} (\bibinfo {year} {2001})}\BibitemShut {NoStop}%
\bibitem [{\citenamefont {Zakharov}\ \emph {et~al.}(2005)\citenamefont
  {Zakharov}, \citenamefont {Korotkevich}, \citenamefont {Pushkarev},\ and\
  \citenamefont {Dyachenko}}]{zakharova2005mesoscopic}%
  \BibitemOpen
  \bibfield  {author} {\bibinfo {author} {\bibfnamefont {V.}~\bibnamefont
  {Zakharov}}, \bibinfo {author} {\bibfnamefont {A.}~\bibnamefont
  {Korotkevich}}, \bibinfo {author} {\bibfnamefont {A.}~\bibnamefont
  {Pushkarev}},\ and\ \bibinfo {author} {\bibfnamefont {A.}~\bibnamefont
  {Dyachenko}},\ }\bibfield  {title} {\bibinfo {title} {Mesoscopic wave
  turbulence},\ }\href@noop {} {\bibfield  {journal} {\bibinfo  {journal} {JETP
  LETTERS}\ }\textbf {\bibinfo {volume} {82}} (\bibinfo {year}
  {2005})}\BibitemShut {NoStop}%
\bibitem [{\citenamefont {Lvov}\ \emph {et~al.}(2006)\citenamefont {Lvov},
  \citenamefont {Nazarenko},\ and\ \citenamefont {Pokorni}}]{LVOV200624}%
  \BibitemOpen
  \bibfield  {author} {\bibinfo {author} {\bibfnamefont {Y.~V.}\ \bibnamefont
  {Lvov}}, \bibinfo {author} {\bibfnamefont {S.}~\bibnamefont {Nazarenko}},\
  and\ \bibinfo {author} {\bibfnamefont {B.}~\bibnamefont {Pokorni}},\
  }\bibfield  {title} {\bibinfo {title} {Discreteness and its effect on
  water-wave turbulence},\ }\href
  {https://doi.org/https://doi.org/10.1016/j.physd.2006.04.003} {\bibfield
  {journal} {\bibinfo  {journal} {Physica D: Nonlinear Phenomena}\ }\textbf
  {\bibinfo {volume} {218}},\ \bibinfo {pages} {24} (\bibinfo {year}
  {2006})}\BibitemShut {NoStop}%
\bibitem [{\citenamefont {Dyachenko}\ \emph {et~al.}(2003)\citenamefont
  {Dyachenko}, \citenamefont {Korotkevich},\ and\ \citenamefont
  {Zakharov}}]{dyachenko2003decay}%
  \BibitemOpen
  \bibfield  {author} {\bibinfo {author} {\bibfnamefont {A.~I.}\ \bibnamefont
  {Dyachenko}}, \bibinfo {author} {\bibfnamefont {A.~O.}\ \bibnamefont
  {Korotkevich}},\ and\ \bibinfo {author} {\bibfnamefont {V.~E.}\ \bibnamefont
  {Zakharov}},\ }\bibfield  {title} {\bibinfo {title} {Decay of the
  monochromatic capillary wave},\ }\href@noop {} {\bibfield  {journal}
  {\bibinfo  {journal} {Journal of Experimental and Theoretical Physics
  Letters}\ }\textbf {\bibinfo {volume} {77}},\ \bibinfo {pages} {477}
  (\bibinfo {year} {2003})}\BibitemShut {NoStop}%
\bibitem [{\citenamefont {Annenkov}\ and\ \citenamefont
  {Shrira}(2006)}]{annenkov2006role}%
  \BibitemOpen
  \bibfield  {author} {\bibinfo {author} {\bibfnamefont {S.~Y.}\ \bibnamefont
  {Annenkov}}\ and\ \bibinfo {author} {\bibfnamefont {V.~I.}\ \bibnamefont
  {Shrira}},\ }\bibfield  {title} {\bibinfo {title} {Role of non-resonant
  interactions in the evolution of nonlinear random water wave fields},\
  }\href@noop {} {\bibfield  {journal} {\bibinfo  {journal} {Journal of Fluid
  Mechanics}\ }\textbf {\bibinfo {volume} {561}},\ \bibinfo {pages} {181}
  (\bibinfo {year} {2006})}\BibitemShut {NoStop}%
\bibitem [{\citenamefont {L'vov}\ and\ \citenamefont
  {Nazarenko}(2010)}]{PhysRevE.82.056322}%
  \BibitemOpen
  \bibfield  {author} {\bibinfo {author} {\bibfnamefont {V.~S.}\ \bibnamefont
  {L'vov}}\ and\ \bibinfo {author} {\bibfnamefont {S.}~\bibnamefont
  {Nazarenko}},\ }\bibfield  {title} {\bibinfo {title} {Discrete and mesoscopic
  regimes of finite-size wave turbulence},\ }\href
  {https://doi.org/10.1103/PhysRevE.82.056322} {\bibfield  {journal} {\bibinfo
  {journal} {Phys. Rev. E}\ }\textbf {\bibinfo {volume} {82}},\ \bibinfo
  {pages} {056322} (\bibinfo {year} {2010})}\BibitemShut {NoStop}%
\bibitem [{\citenamefont {Shukla}\ and\ \citenamefont
  {Nazarenko}(2021)}]{vish}%
  \BibitemOpen
  \bibfield  {author} {\bibinfo {author} {\bibfnamefont {V.}~\bibnamefont
  {Shukla}}\ and\ \bibinfo {author} {\bibfnamefont {S.}~\bibnamefont
  {Nazarenko}},\ }\bibfield  {title} {\bibinfo {title} {Non-equilibrium
  bose-einstein condensation},\ }\href@noop {} {\bibfield  {journal} {\bibinfo
  {journal} {arXiv preprint arXiv:2105.07274}\ } (\bibinfo {year}
  {2021})}\BibitemShut {NoStop}%
\bibitem [{\citenamefont {Connaughton}\ \emph {et~al.}(2005)\citenamefont
  {Connaughton}, \citenamefont {Josserand}, \citenamefont {Picozzi},
  \citenamefont {Pomeau},\ and\ \citenamefont {Rica}}]{Connaughton2005}%
  \BibitemOpen
  \bibfield  {author} {\bibinfo {author} {\bibfnamefont {C.}~\bibnamefont
  {Connaughton}}, \bibinfo {author} {\bibfnamefont {C.}~\bibnamefont
  {Josserand}}, \bibinfo {author} {\bibfnamefont {A.}~\bibnamefont {Picozzi}},
  \bibinfo {author} {\bibfnamefont {Y.}~\bibnamefont {Pomeau}},\ and\ \bibinfo
  {author} {\bibfnamefont {S.}~\bibnamefont {Rica}},\ }\bibfield  {title}
  {\bibinfo {title} {Condensation of classical nonlinear waves},\ }\href
  {https://doi.org/10.1103/PhysRevLett.95.263901} {\bibfield  {journal}
  {\bibinfo  {journal} {Phys. Rev. Lett.}\ }\textbf {\bibinfo {volume} {95}},\
  \bibinfo {pages} {263901} (\bibinfo {year} {2005})}\BibitemShut {NoStop}%
\bibitem [{\citenamefont {Falkovich}(1994)}]{Falkovich}%
  \BibitemOpen
  \bibfield  {author} {\bibinfo {author} {\bibfnamefont {G.}~\bibnamefont
  {Falkovich}},\ }\bibfield  {title} {\bibinfo {title} {Bottleneck phenomenon
  in developed turbulence},\ }\href {https://doi.org/10.1063/1.868255}
  {\bibfield  {journal} {\bibinfo  {journal} {Physics of Fluids}\ }\textbf
  {\bibinfo {volume} {6}},\ \bibinfo {pages} {1411} (\bibinfo {year} {1994})},\
  \Eprint {https://arxiv.org/abs/https://doi.org/10.1063/1.868255}
  {https://doi.org/10.1063/1.868255} \BibitemShut {NoStop}%
\bibitem [{\citenamefont {Borue}\ and\ \citenamefont
  {Orszag}(1995)}]{borue1995self}%
  \BibitemOpen
  \bibfield  {author} {\bibinfo {author} {\bibfnamefont {V.}~\bibnamefont
  {Borue}}\ and\ \bibinfo {author} {\bibfnamefont {S.~A.}\ \bibnamefont
  {Orszag}},\ }\bibfield  {title} {\bibinfo {title} {Self-similar decay of
  three-dimensional homogeneous turbulence with hyperviscosity},\ }\href@noop
  {} {\bibfield  {journal} {\bibinfo  {journal} {Physical Review E}\ }\textbf
  {\bibinfo {volume} {51}},\ \bibinfo {pages} {R856} (\bibinfo {year}
  {1995})}\BibitemShut {NoStop}%
\bibitem [{\citenamefont {Cichowlas}\ \emph {et~al.}(2005)\citenamefont
  {Cichowlas}, \citenamefont {Bona\"{\i}ti}, \citenamefont {Debbasch},\ and\
  \citenamefont {Brachet}}]{Cichowlas}%
  \BibitemOpen
  \bibfield  {author} {\bibinfo {author} {\bibfnamefont {C.}~\bibnamefont
  {Cichowlas}}, \bibinfo {author} {\bibfnamefont {P.}~\bibnamefont
  {Bona\"{\i}ti}}, \bibinfo {author} {\bibfnamefont {F.}~\bibnamefont
  {Debbasch}},\ and\ \bibinfo {author} {\bibfnamefont {M.}~\bibnamefont
  {Brachet}},\ }\bibfield  {title} {\bibinfo {title} {Effective dissipation and
  turbulence in spectrally truncated euler flows},\ }\href
  {https://doi.org/10.1103/PhysRevLett.95.264502} {\bibfield  {journal}
  {\bibinfo  {journal} {Phys. Rev. Lett.}\ }\textbf {\bibinfo {volume} {95}},\
  \bibinfo {pages} {264502} (\bibinfo {year} {2005})}\BibitemShut {NoStop}%
\bibitem [{\citenamefont {Krstulovic}\ and\ \citenamefont
  {Brachet}(2011)}]{Dispersive}%
  \BibitemOpen
  \bibfield  {author} {\bibinfo {author} {\bibfnamefont {G.}~\bibnamefont
  {Krstulovic}}\ and\ \bibinfo {author} {\bibfnamefont {M.}~\bibnamefont
  {Brachet}},\ }\bibfield  {title} {\bibinfo {title} {Dispersive bottleneck
  delaying thermalization of turbulent bose-einstein condensates},\ }\href
  {https://doi.org/10.1103/PhysRevLett.106.115303} {\bibfield  {journal}
  {\bibinfo  {journal} {Phys. Rev. Lett.}\ }\textbf {\bibinfo {volume} {106}},\
  \bibinfo {pages} {115303} (\bibinfo {year} {2011})}\BibitemShut {NoStop}%
\bibitem [{\citenamefont {Chibbaro}\ \emph {et~al.}(2017)\citenamefont
  {Chibbaro}, \citenamefont {Dematteis}, \citenamefont {Josserand},\ and\
  \citenamefont {Rondoni}}]{Chibbaro}%
  \BibitemOpen
  \bibfield  {author} {\bibinfo {author} {\bibfnamefont {S.}~\bibnamefont
  {Chibbaro}}, \bibinfo {author} {\bibfnamefont {G.}~\bibnamefont {Dematteis}},
  \bibinfo {author} {\bibfnamefont {C.}~\bibnamefont {Josserand}},\ and\
  \bibinfo {author} {\bibfnamefont {L.}~\bibnamefont {Rondoni}},\ }\bibfield
  {title} {\bibinfo {title} {Wave-turbulence theory of four-wave nonlinear
  interactions},\ }\href {https://doi.org/10.1103/PhysRevE.96.021101}
  {\bibfield  {journal} {\bibinfo  {journal} {Phys. Rev. E}\ }\textbf {\bibinfo
  {volume} {96}},\ \bibinfo {pages} {021101(R)} (\bibinfo {year}
  {2017})}\BibitemShut {NoStop}%
\bibitem [{\citenamefont {Fujimoto}\ and\ \citenamefont
  {Tsubota}(2015)}]{fujimoto2015bogoliubov}%
  \BibitemOpen
  \bibfield  {author} {\bibinfo {author} {\bibfnamefont {K.}~\bibnamefont
  {Fujimoto}}\ and\ \bibinfo {author} {\bibfnamefont {M.}~\bibnamefont
  {Tsubota}},\ }\bibfield  {title} {\bibinfo {title} {Bogoliubov-wave
  turbulence in bose-einstein condensates},\ }\href@noop {} {\bibfield
  {journal} {\bibinfo  {journal} {Physical Review A}\ }\textbf {\bibinfo
  {volume} {91}},\ \bibinfo {pages} {053620} (\bibinfo {year}
  {2015})}\BibitemShut {NoStop}%
\bibitem [{\citenamefont {Hossain}\ and\ \citenamefont
  {Islam}(2014)}]{Hossain}%
  \BibitemOpen
  \bibfield  {author} {\bibinfo {author} {\bibfnamefont {M.~A.}\ \bibnamefont
  {Hossain}}\ and\ \bibinfo {author} {\bibfnamefont {M.~S.}\ \bibnamefont
  {Islam}},\ }\bibfield  {title} {\bibinfo {title} {{Generalized composite
  numerical integration rule over a polygon using Gaussian quadrature}},\
  }\href@noop {} {\bibfield  {journal} {\bibinfo  {journal} {Dhaka University
  Journal of Science}\ }\textbf {\bibinfo {volume} {62}},\ \bibinfo {pages}
  {25} (\bibinfo {year} {2014})}\BibitemShut {NoStop}%
\bibitem [{\citenamefont {Salzer}(1972)}]{bar}%
  \BibitemOpen
  \bibfield  {author} {\bibinfo {author} {\bibfnamefont {H.~E.}\ \bibnamefont
  {Salzer}},\ }\bibfield  {title} {\bibinfo {title} {{Lagrangian Interpolation
  at the Chebyshev Points $x_{n,\nu}=\cos(\nu\pi/n)$, $\nu=O(1)n$; some Unnoted
  Advantages}},\ }\href {https://doi.org/10.1093/comjnl/15.2.156} {\bibfield
  {journal} {\bibinfo  {journal} {The Computer Journal}\ }\textbf {\bibinfo
  {volume} {15}},\ \bibinfo {pages} {156} (\bibinfo {year} {1972})}\BibitemShut
  {NoStop}%
\bibitem [{\citenamefont {Baltensperger}\ \emph {et~al.}(1999)\citenamefont
  {Baltensperger}, \citenamefont {Berrut},\ and\ \citenamefont
  {No{\"{e}}l}}]{Bur}%
  \BibitemOpen
  \bibfield  {author} {\bibinfo {author} {\bibfnamefont {R.}~\bibnamefont
  {Baltensperger}}, \bibinfo {author} {\bibfnamefont {J.}~\bibnamefont
  {Berrut}},\ and\ \bibinfo {author} {\bibfnamefont {B.}~\bibnamefont
  {No{\"{e}}l}},\ }\bibfield  {title} {\bibinfo {title} {{Exponential
  convergence of a linear rational interpolant between transformed Chebyshev
  points}},\ }\href {https://doi.org/10.1090/S0025-5718-99-01070-4} {\bibfield
  {journal} {\bibinfo  {journal} {Math. Comput.}\ }\textbf {\bibinfo {volume}
  {68}},\ \bibinfo {pages} {1109} (\bibinfo {year} {1999})}\BibitemShut
  {NoStop}%
\bibitem [{\citenamefont {Tee}\ and\ \citenamefont
  {Trefethen}(2006)}]{TeeTref}%
  \BibitemOpen
  \bibfield  {author} {\bibinfo {author} {\bibfnamefont {T.~W.}\ \bibnamefont
  {Tee}}\ and\ \bibinfo {author} {\bibfnamefont {L.~N.}\ \bibnamefont
  {Trefethen}},\ }\bibfield  {title} {\bibinfo {title} {{A Rational Spectral
  Collocation Method with Adaptively Transformed Chebyshev Grid Points}},\
  }\href {https://doi.org/10.1137/050641296} {\bibfield  {journal} {\bibinfo
  {journal} {SIAM Journal on Scientific Computing}\ }\textbf {\bibinfo {volume}
  {28}},\ \bibinfo {pages} {1798} (\bibinfo {year} {2006})},\ \Eprint
  {https://arxiv.org/abs/https://doi.org/10.1137/050641296}
  {https://doi.org/10.1137/050641296} \BibitemShut {NoStop}%
\bibitem [{\citenamefont {Jafari-Varzaneh}\ and\ \citenamefont
  {Hosseini}(2015)}]{Hoss}%
  \BibitemOpen
  \bibfield  {author} {\bibinfo {author} {\bibfnamefont {H.~A.}\ \bibnamefont
  {Jafari-Varzaneh}}\ and\ \bibinfo {author} {\bibfnamefont {S.~M.}\
  \bibnamefont {Hosseini}},\ }\bibfield  {title} {\bibinfo {title} {{A new map
  for the Chebyshev pseudospectral solution of differential equations with
  large gradients}},\ }\href@noop {} {\bibfield  {journal} {\bibinfo  {journal}
  {Numerical Algorithms}\ }\textbf {\bibinfo {volume} {69}},\ \bibinfo {pages}
  {95} (\bibinfo {year} {2015})}\BibitemShut {NoStop}%
\bibitem [{\citenamefont {Semisalov}\ and\ \citenamefont
  {Kuzmin}(2017)}]{SemKuz}%
  \BibitemOpen
  \bibfield  {author} {\bibinfo {author} {\bibfnamefont {B.}~\bibnamefont
  {Semisalov}}\ and\ \bibinfo {author} {\bibfnamefont {G.}~\bibnamefont
  {Kuzmin}},\ }\bibfield  {title} {\bibinfo {title} {{Modification of Fourier
  Approximation for Solving Boundary Value Problems Having Singularities of
  Boundary Layer Type}}\ }(\bibinfo {year} {2017})\ pp.\ \bibinfo {pages}
  {406--422}\BibitemShut {NoStop}%
\bibitem [{\citenamefont {Idimeshev}(2020)}]{Semen}%
  \BibitemOpen
  \bibfield  {author} {\bibinfo {author} {\bibfnamefont {S.}~\bibnamefont
  {Idimeshev}},\ }\bibfield  {title} {\bibinfo {title} {{Rational approximation
  in initial boundary value problems with fronts (in Russian)}},\ }\href@noop
  {} {\bibfield  {journal} {\bibinfo  {journal} {Computational technologies}\
  }\textbf {\bibinfo {volume} {25}},\ \bibinfo {pages} {69} (\bibinfo {year}
  {2020})}\BibitemShut {NoStop}%
\bibitem [{\citenamefont {Gentleman}(1972)}]{Clenshaw}%
  \BibitemOpen
  \bibfield  {author} {\bibinfo {author} {\bibfnamefont {W.~M.}\ \bibnamefont
  {Gentleman}},\ }\bibfield  {title} {\bibinfo {title} {{Implementing
  Clenshaw--Curtis quadrature, II computing the cosine transformation}},\
  }\href@noop {} {\bibfield  {journal} {\bibinfo  {journal} {Communications of
  the ACM}\ }\textbf {\bibinfo {volume} {15}},\ \bibinfo {pages} {343}
  (\bibinfo {year} {1972})}\BibitemShut {NoStop}%
\end{thebibliography}%

\end{document}
%